\documentclass[fleqn,usenatbib]{mnras}

\usepackage{newtxtext,newtxmath}
\usepackage[version=4]{mhchem}

\usepackage[T1]{fontenc}

\DeclareRobustCommand{\VAN}[3]{#2}
\let\VANthebibliography\thebibliography
\def\thebibliography{\DeclareRobustCommand{\VAN}[3]{##3}\VANthebibliography}

\usepackage{graphicx}	
\usepackage{amsmath}	
\usepackage{lineno}
\usepackage{CJKutf8}
\defcitealias{Liu_2025_PartI}{Part~I} 

\title[Transport-induced chemistry on K2-18b, Part 2]{Three-dimensional transport-induced chemistry on
temperate sub-Neptune K2-18b, Part II: the combined effects of atmospheric dynamics and chemical reactions}

\author[J. Liu et al.]{
\begin{CJK*}{UTF8}{gbsn}Jiachen Liu (刘伽晨),$^{1,2,3}$\thanks{E-mail: jiachenliu@mpia.de}
Duncan Christie,$^{1,3}$
Jun Yang (杨军)$^{2}$\end{CJK*}, Krisztian Kohary$^{3}$
\\
$^{1}$Max Planck Institute for Astronomy, Heidelberg, 69117, Germany\\
$^{2}$Department of Atmospheric and Oceanic Sciences, School of Physics, Peking University, Beijing 100871, People's Republic of China\\
$^{3}$Department of Physics and Astronomy, Faculty of Environment, Science and Economy, University of Exeter, Exeter EX4 4QL, UK
}

\date{Accepted XXX. Received YYY; in original form ZZZ}

\pubyear{\the\year{}}

\begin{document}
\label{firstpage}
\pagerange{\pageref{firstpage}--\pageref{lastpage}}
\maketitle

\begin{abstract}
The upper atmospheres of temperate sub-Neptunes are strongly influenced by atmospheric dynamics due to their cool equilibrium temperature and thereby longer chemical timescales than the atmospheric dynamical timescales. In this study, we used a three-dimensional (3D) general circulation model to investigate the transport-induced disequilibrium chemistry and vertical mixing on temperate gas-rich mini-Neptunes, using K2-18b as an example. We model K2-18b assuming 180 times solar metallicity and consider it as either a synchronous or an asynchronous rotator, exploring spin-orbit resonances of 2:1, 6:1, and 10:1. We find that the vertical transport affects the chemical structure significantly, making CO$_2$ and CO more abundant ($\sim$10$^{-3}$) in the upper atmosphere compared to the chemical equilibrium abundance (\textless{10$^{-15}$}), and horizontal winds further homogenize the chemical composition zonally in this region.  Molecular abundances in the photosphere generally agree across different rotation periods. We employ a passive tracer in the model to estimate the one-dimensional (1D) equivalent eddy-diffusion coefficient ($K_{zz}$) of K2-18b, providing a parameter useful for future 1D atmospheric models. Additionally, synthetic transmission spectra generated from our model are compared with the JWST observations, and we find that our model can provide a comparable fit to the observations. This work offers a 3D perspective on transport-induced chemistry on a temperate sub-Neptune and derives vertical mixing parameters to support 1D modelling.
\end{abstract}




\section{Introduction} \label{sec:intro}

Sub-Neptunes, defined as planets with radii between 1.7 and 3.5 R$_\oplus$, are among the most abundant class of (exo)planets detected to date \citep{batalha2014exploring,fulton2018california}.
However, the interior structure of these planets remains poorly constrained. In particular, their bulk densities permit a wide range of degenerate interior and atmospheric compositions, encompassing varying proportions of rock, H$_2$-He gas, volatiles (e.g. H$_2$O), and ices \citep[e.g.][]{rogers2010framework,rogers2023conclusive}. One prevailing interpretation is that they are scaled-down analogues of gas giants, consisting of rocky cores enveloped by primordial H$_2$/He atmospheres \citep[e.g.][]{misener2021cool}. Alternatively, they may represent `water worlds', characterised by substantial H$_2$O-rich atmospheres, ices, and possibly liquid layers \citep[e.g.][]{Zeng_planet_distribution_2019,luque_density_2022}.
The debate is particularly active for temperate sub-Neptunes, which receive stellar fluxes comparable to that of Earth. Some studies have proposed that these planets could host liquid oceans beneath their atmospheres, classifying them as `Hycean worlds' \citep{madhusudhan_interior_2020,Piette_Madhu_2020,Nixon_Madhu_ocean_2021}. 
Potential clear-sky conditions on temperate sub-Neptunes make them promising targets for molecular detections with JWST, providing an observational window into breaking the degeneracy in their structural interpretations \citep{yu2021haze,dymont2022cleaning,brande_clouds_2024}. 

One temperate sub-Neptune that has been characterised and has molecules detected by JWST is K2-18b. It has a mass and radius of approximately 8.6~M$_\oplus$ and 2.6~R$_\oplus$ and an orbital period of 32.94 Earth days\footnote{Unless otherwise specified, `day' hereafter refers to an Earth day.} \citep{Montet_K2_2015,benneke_water_2019}. K2-18b receives a stellar flux of $\sim$1367 W\,m$^{-2}$ and has an equilibrium temperature of $\sim$280~K with zero planetary albedo. \citet{madhusudhan_carbon-bearing_2023} presented the first JWST transmission spectrum of K2-18b, which reported the detection of $\sim$1\% volume mixing ratios of both CH$_4$ and CO$_2$. This result suggests a potentially H$_2$-rich atmosphere with abundant carbon-bearing species, motivating further exploration of both its atmospheric chemistry and interior structure. However, the interpretation of these findings remains debated.

Rather than resolving the structural degeneracy, this JWST observation of K2-18b has prompted further scientific debate regarding its nature. \citet{madhusudhan_carbon-bearing_2023} argued that the detection of CH$_4$ and CO$_2$ and the non-detection of NH$_3$ supports a Hycean world scenario, as NH$_3$ depletion is indicative of a liquid water ocean \citep{Hu_solubility_nh3_2021}. This interpretation was later supported by \citet{cooke2024considerations} but was challenged by \citet{shorttle_distinguishing_2024} and \citet{wogan2024jwst}. \citet{shorttle_distinguishing_2024} proposed that a surface magma ocean can also solubilise NH$_3$, and \citet{wogan2024jwst} suggested that a gas-rich mini-Neptune\footnote{While `sub-Neptune' classifies planets based on size (analogous to super-Earths), `mini-Neptune' refers to classification based on interior structure (analogous to Hycean worlds). For a thorough discussion of sub-Neptunes' interior and envelope structures, please see \citet{benneke2024jwst}.} can adequately explain the observation. Subsequently, \citet{schmidt2025comprehensive} argued that the CO$_2$ detection in \citet{madhusudhan_carbon-bearing_2023} is not statistically significant and suggested that K2-18b can instead be explained by an oxygen-poor mini-Neptune scenario. Follow-up observations by \citet{Hu_K2-18b_2025}, however, confirmed the detections of CH$_4$ and CO$_2$ and the non-detections of CO and NH$_3$.

Moreover, the claim of a possible biosignature gas detection in \citet{madhusudhan_carbon-bearing_2023} and \citet{madhusudhan2025new} has generated considerable debate regarding the true nature of K2-18b \citep{seager_prospects_nodate,taylor_are_2025,welbanks_challenges_2025,luque_insufficient_2025,Stevenson_2025_K2-18b}. In parallel, climate modelling studies have also raised concerns about the viability of the Hycean world scenario for K2-18b \citep{pierrehumbert_runaway_2023,innes2023runaway}. The intense greenhouse effect driven by its H$_2$-rich atmosphere may push the planet into a runaway greenhouse state, potentially resulting in a supercritical phase \citep{Pierrehumbert_2011,koll_hot_2019,pierrehumbert_runaway_2023,innes2023runaway}.

The degeneracy in K2-18b's interior structure is unlikely to be resolved without both improved observational constraints and more comprehensive modelling of the various proposed scenarios. %
In this study, we focus on the mini-Neptune interpretation of K2-18b. A more detailed understanding of the planet as a gas-rich mini-Neptune may offer critical insights into resolving this structural degeneracy. We note that this study was initiated at a time when the observational data and analysis from \citet{madhusudhan_carbon-bearing_2023} served as the primary reference available.  
The transmission spectra reported by \citet{Hu_K2-18b_2025} were published while we were analysing our simulation results. Although our simulation setup did not incorporate these new observations, we compared our results with their data for completeness. JWST has observed additional transmission spectra of K2-18b but has not yet been released, which will provide further insights into the planet’s nature \citep{Hu_JWST_2021}.

The chemical composition of mini-Neptune atmospheres is primarily determined by thermochemical kinetics, vertical transport, and photochemistry. In the hot deep atmosphere, chemical timescales ($\tau_\mathrm{chem}$) are short, so the composition is close to chemical equilibrium. Moving to cooler, higher altitudes, vertical transport can dominate when the chemical timescale exceeds the vertical mixing timescale ($\tau_\mathrm{mix}$). The level where $\tau_\mathrm{chem} = \tau_\mathrm{mix}$ is generally referred to as the quench level, above which chemical abundances deviate from equilibrium values due to atmospheric transport. In the upper atmosphere, where optical depth decreases and UV fluxes become significant, photochemistry alters the abundances of chemical species. For K2-18b, the cool upper atmosphere leads to extremely long chemical timescales ($\gg 10^{10}$ s), so vertical transport strongly influences the observed chemical composition \citep[][fig. 3]{Liu_2025_PartI}. These variations in chemical abundances can have a substantial impact on the transmission spectra.

Previous studies on the transport-induced disequilibrium chemistry of K2-18b as a mini-Neptune relied on one-dimensional (1D) models \citep[e.g.][]{blain_1d_2021,tsai_inferring_2021,yu__how_2021,bezard_methane_2022,wogan2024jwst,cooke2024considerations}. In these models, the vertical mixing process is parametrized by the eddy diffusion coefficient ($K_{zz}$). In full chemical kinetics–transport models, $K_{zz}$ is included as an important parameter in the continuity equations of chemical species \citep{Gladstone_1996,tsai_inferring_2021,yu__how_2021,wogan2024jwst,cooke2024considerations}, whereas in quenching approximation models \citep{blain_1d_2021,bezard_methane_2022}, the abundances of species are assumed to freeze above the quench level—where $\tau_\mathrm{mix}$ ($H^2/K_{zz}$, in which $H$ is the atmospheric scale height) equals $\tau_\mathrm{chem}$. However, the $K_{zz}$ values used in these studies remain coarsely constrained. \citet{blain_1d_2021} and \citet{bezard_methane_2022} assumed constant $K_{zz}$ values in their simulations. Although \citet{tsai_inferring_2021}, \citet{yu__how_2021}, \citet{wogan2024jwst}, and \citet{cooke2024considerations} employed 1D $K_{zz}$ profiles, these $K_{zz}$ profiles were estimated based on empirical formulas \citep{yu__how_2021,wogan2024jwst,cooke2024considerations}, or derived from previous three-dimensional (3D) general circulation model (GCM) simulations of other (not K2-18b) gas-rich planets \citep{tsai_inferring_2021}. The coarse constraint of $K_{zz}$ makes the results from 1D models less reliable.

Moreover, while 1D models provide useful insights, they cannot capture the inherently 3D nature of exoplanetary atmospheres. When $\tau_\mathrm{chem} \gg \tau_\mathrm{adv}$ (the horizontal advection timescale), atmospheric dynamics can significantly influence the horizontal distribution of chemical species. Therefore, 1D models neglecting horizontal transport are limited in their ability to fully represent 3D transport-induced chemistry on K2-18b.

Several previous studies have performed 3D cloud-free and haze-free coupled chemistry-radiation-hydrodynamics simulations for the hot gas giant planets: hot Jupiters HD 189733b, HD 209458b, WASP-17b, and WASP-96b, and a warm Neptune, HAT-P-11b \citep{drummond2020implications,zamyatina_observability_2022,zamyatina_quenching-driven_2024}. These simulations have demonstrated that atmospheric circulation can drive substantial departures from local chemical equilibrium, leading to observable differences in predicted spectra. For instance, \citet{drummond2020implications} found that the 3D advection substantially alters the chemical composition of hot Jupiters HD 209458b and HD 189733b, compared with a simulation assuming local chemical equilibrium. These compositional changes, particularly in CH$_4$, CO, and NH$_3$, lead to significant changes in synthetic transmission and emission spectra.

However, such coupled chemistry-radiation-hydrodynamics simulations have not yet been applied to cooler and smaller planets, such as temperate mini-Neptunes. Modelling these planets presents challenges. First, due to slower chemical reaction rates at lower temperatures, much of their atmosphere is governed by transport-induced disequilibrium chemistry. This shifts the quench level deeper into the atmosphere, requiring not only the observable photosphere (typically between $\sim$10$^{-2}$ and $\sim$10$^{-4}$ bar) but also deeper layers to reach dynamical and chemical convergence. Achieving this demands long simulation times, which impose stricter demands on the model's numerical stability and conservation properties, while also substantially increasing the computational cost.
Secondly, given temperate mini-Neptunes' greater orbital distance, their tidal synchronisation timescales may be long, and they may not be in synchronous rotation \citep{guillot1996giant}. This possibility requires the exploration of asynchronous rotation states in 3D modelling. 

In this study, we provide a 3D perspective on transport-induced disequilibrium chemistry on temperate mini-Neptune K2-18b with a consistently coupled chemistry-radiation-hydrodynamics model. This paper is Part II of a series. 
In Part I of the study \citep{Liu_2025_PartI}, we characterise the atmospheric circulation and analyse its influence on 3D transport. We model K2-18b as a synchronous rotator or as an asynchronous rotator with spin-orbit resonance (SOR) of 2:1, 6:1, and 10:1, where the ratio denotes the number of planetary rotations per orbit. We restrict our parameter space to SOR $\geq 1$ because gas-rich planets are generally expected to form with relatively rapid rotation and then tidally spin down toward synchronous rotation \citep[e.g.][]{Batygin_2018_giant,Bryan_spin_2018,Mathis_tidal_2018}. Although sub-synchronous rotation (SOR $<1$) can in principle arise through companion-driven secular spin-orbit interactions in multiplanet systems \citep[e.g.][]{Millholland_2018,Su_2022,Lai_2026}, such a scenario is not strongly motivated for K2-18b, since the proposed inner companion K2-18c remains unconfirmed \citep{Cloutier_2017,Sarkis_2018,Cloutier_2019}. Moreover, for a 33-day orbit, the synchronous case already lies in the slow-rotation regime \citepalias[][equation 5]{Liu_2025_PartI}, so reducing the rotation rate further is not expected to qualitatively alter the large-scale circulation. To trace atmospheric transport, we employed a passive tracer, which is advected by the flow but does not interact with radiation or participate in chemical reactions. A brief review of the main results in \citetalias{Liu_2025_PartI} is presented in section~\ref{subsec:overview} below. Although we do not analyse the chemical structure in \citetalias{Liu_2025_PartI}, we emphasise that those simulations do incorporate full chemical kinetics. 

Before \citetalias{Liu_2025_PartI}, although three studies have presented 3D simulations of K2-18b as gas-rich mini-Neptune \citep{charnay2021formation,innes2022atmospheric,Barrier_convection_2025}, none incorporate transport-induced disequilibrium chemistry, instead focusing on other aspects of K2-18b’s atmosphere. \citet{charnay2021formation} examined water cloud formation and its observational impact, \citet{innes2022atmospheric} investigated the dry atmospheric circulation, and \citet{Barrier_convection_2025} employed a new convection scheme to model convection on K2-18b.

In this paper, we extend the analysis to active chemical species and reactions, excluding photochemistry. We compare the 3D simulation results to those from a 1D chemical kinetics–transport model and derive an equivalent $K_{zz}$ profile suitable for further use in 1D models of K2-18b. Synthetic transmission spectra generated from our 3D models are also compared with the JWST observations in \citet{madhusudhan_carbon-bearing_2023} and \citet{Hu_K2-18b_2025}.

The structure of this paper is as follows: section~\ref{sec:methods} introduces the GCM we used in this study and the experimental design. Section~\ref{subsec:overview} reviews the thermal structure, atmospheric circulation, and the passive tracer distribution in the 3D simulations, as previously analysed in \citetalias{Liu_2025_PartI}. In section~\ref{subsec:chemical}, we present the 3D chemical structure. The derived $K_{zz}$ profile and the comparison between results from the 1D and 3D models are shown in section~\ref{subsec:Kzz}. Section~\ref{subsec:observation} compares our synthetic spectra with JWST observations. Conclusions and discussion are given in section~\ref{sec:conclusions}.

\section{Model Description and Experimental Design} \label{sec:methods}

\begin{table}
	\centering
	\caption{Parameters used in this study.}
	\label{k2-18b}
	\begin{tabular}{cc} 
		\hline
		\textbf{Parameter} & \textbf{K2-18b}\\
		\hline
		Stellar flux & 1367~W\,m$^{-2}$  \\
            Inner radius ($R_\mathrm{p}$)& 16,430~km \\
            Orbital period & 32.94~days \\
            Planetary mass & 5.15$\times$10$^{25}$~kg \\
            Semi-major axis & 0.1591~au \\
            Surface gravity ($g_\mathrm{p}$) & 12.44~m\,s$^{-2}$ \\  
            Atmospheric composition & 180 $\times$ solar metallicity\\
            Specific gas constant ($R$)& 1023~J\,kg$^{-1}$\,K$^{-1}$ \\
            Specific heat capacity ($c_p$) & 3733~J\,kg$^{-1}$\,K$^{-1}$ \\
            Mean molecular weight & 8.46~g\,mol$^{-1}$ \\
            Internal temperature & 60~K (3.7~W\,m$^{-2}$) \\
            Domain height & 800~km \\
            Bottom and upper pressure & 200 and $\sim$10$^{-5}$~bar \\
            Horizontal resolution & 2.5$^{\circ}$$\times$2$^{\circ}$\\
		\hline
	\end{tabular}
\end{table}

In this study, we simulate the atmosphere of K2-18b using the Met Office 3D Unified Model (UM) GCM. The main parameters used in this study for K2-18b can be found in Table \ref{k2-18b}.
The UM has been used to simulate hydrogen-dominated hot Jupiters \citep[e.g.][]{mayne2014unified,amundsen2016uk,drummond_effects_2016,mayne2017results,drummond_observable_2018,drummond2020implications,zamyatina_observability_2022,christie_impact_2021} and sub-Neptunes \citep[e.g.][]{drummond_effect_2018,mayne2019limits,christie_impact_2022}. It solves the non-hydrostatic equations of motion using a finite difference semi-implicit semi-Lagrangian scheme on an Arakawa C grid \citep{wood2014inherently}. A thorough description of the model set-up for the hydrogen-dominated atmosphere can be found in \citet{mayne2014unified}.
The model employs a geometric height-based vertical grid. Gravity is allowed to vary with height, following $g(r) = g_\mathrm{p}(R_\mathrm{p}/r)^2$, where $g_\mathrm{p}$ and $R_\mathrm{p}$ denote the surface gravity and planetary inner radius, respectively, and $r$ is the radial distance from the planet's center.
The horizontal resolution is 2.5$^{\circ}$$\times$2$^{\circ}$ in longitude and latitude. The model domain extends up to 800~km with 66 levels evenly spaced in height, spanning pressures from 200~bar down to approximately 1~Pa.

The radiative scheme used in the model is the two-stream correlated-k radiative transfer model SOCRATES \citep{edwards1996studies,edwards1996efficient}, with 32 spectral bands covering 0.2-322~$\mu$m. The opacity of H$_2$O, CH$_4$, NH$_3$, CO$_2$, and CO, the H$_2$-H$_2$ collision-induced absorption, H$_2$-He collision-induced absorption, and the Rayleigh scattering due to H$_2$ and He are included in this study. We use the stellar spectrum of GJ 176, an M2.5 star with an effective temperature of 3,670 K, measured by the MUSCLES survey \citep{france2016muscles}, as the K2-18b's spectrum has not been measured yet. For all the simulations, we set the internal temperature to 60 K, corresponding to an internal heat source of 3.7~W\,m$^{-2}$, following \citet{charnay2021formation}. 

In \citetalias{Liu_2025_PartI} of this study, we argue that K2-18b may not have enough time to be fully 1:1 tidally locked, due to the comparable magnitudes of K2-18b's tidal spin-down timescale ($\sim$1 Gyr) and the estimated timescale of the K2-18 system ($\sim$2.4 Gyr). Thus, besides simulating K2-18b as a synchronous rotator, we explore its SOR of 2:1, 6:1, and 10:1, corresponding to rotation periods of 16.47, 5.49, and 3.29~days, respectively. The obliquity and eccentricity are set to zero for simplicity.

Given the potentially long convergence times in 3D simulations of sub-Neptunes and the substantial computational time of the chemical kinetics scheme, the 3D simulations are conducted in two steps. First, we spin up the model by prescribing opacity (referred to as the `fixed abundance run') until the total kinetic energy stabilises or the energy imbalance at the top of the atmosphere is limited to $\pm$\,3~W\,m$^{-2}$. The fixed abundance runs are initialised at rest, with a globally uniform temperature profile from the 1D radiative-convective-equilibrium model \texttt{ATMO} \citep{tremblin_fingering_2015,tremblin2016cloudless,drummond_effects_2016}. The model assumes chemical equilibrium, stellar flux uniformly distributed across the planetary surface, a solar C/O ratio (0.55), and 180 times solar metallicity. The abundances of the chemical species that affect opacity (i.e., H$
_2$O, CH$_4$, NH$_3$, CO$_2$, CO, H$_2$, and He) are set according to the vertical average value derived from the chemical kinetics–transport simulations using \texttt{ATMO}, using the same assumptions as above, as well as 10$^{6}$~cm$^2$\,s$^{-1}$ $K_{zz}$. The specific heat capacity, gas constant, and mean molecular weight are derived from the vertically averaged values in the same 1D chemical kinetics–transport simulation (Table \ref{k2-18b}). The parameters used in the 1D model, including the 180 times solar metallicity and $K_{zz}$ values, are chosen based on the best agreement between the synthetic transmission spectrum generated from the 1D model and the transmission spectrum in \citet[][fig. 3]{madhusudhan_carbon-bearing_2023}.
A more detailed discussion of the parameters can be found in Appendix C of \citetalias{Liu_2025_PartI}. The fixed abundance runs are run for 5500 days for synchronous and 2:1 SOR simulations, 8000 and 10000 days for 6:1 SOR and 10:1 SOR simulations, respectively.

After the fixed abundance runs, the simulations are restarted with the coupled chemical kinetics scheme (`kinetics run'). 
This chemical kinetics scheme has been used to study the transport-induced disequilibrium chemistry in hot gas giant atmospheres \citep{drummond2020implications,zamyatina_observability_2022,zamyatina_quenching-driven_2024}. It solves ordinary differential equations (ODEs) to describe the production and loss of the chemical species based on a chosen chemical network using the DLSODES solver \citep{hindmarsh1983scientific}. The chemical network employed in the model is developed by \cite{venot2019reduced}, with 30 chemical species and 181 reversible reactions, excluding photodissociation reactions. The chemical species contained in the scheme are CO, CO$_2$, CH$_4$, NH$_3$, H$_2$O, H$_2$, He, HCN, H, OH, N$_2$, CH$_3$, H$_2$CO, O($^3$P), NH$_2$, HCO, NH, CH$_3$OH, CH$_3$O, NCO, CH$_2$OH, N$_2$H$_2$, NNH, CN, HNCO, $^3$CH$_2$, N$_2$H$_3$, $^1$CH$_2$, HOCN, and H$_2$CN. This network is reduced from \cite{venot_chemical_2012}, which contains 105 species and about 1000 reversible reactions. The reduced network is designed to remove certain chemical species and reactions while preserving accuracy for the observable species H$_2$O, CH$_4$, CO, CO$_2$, NH$_3$, and HCN. It has been validated to closely match the original network across a wide range of temperatures, pressures, metallicities, and C/O ratios \citep{venot2019reduced}.
Each chemical species is treated as a tracer and then advected by the dynamical core in units of mass mixing ratio. For simplicity, we do not include the condensation processes of chemical species in the model. A more detailed description of the chemical kinetics scheme can be found in \cite{drummond2020implications}.

The kinetics runs are initialised with the 3D temperature and wind fields from the end of the fixed abundance runs. Assuming 180 times solar metallicity and solar C/O ratio, the chemical abundances are initialised to their chemical equilibrium values using the Gibbs energy minimisation \citep{drummond_effects_2016,drummond_effect_2018} calculated by the initial 3D temperature fields. The specific gas constant, specific heat capacity, and mean molecular weight are the same as in the fixed abundance runs (Table \ref{k2-18b}).

We also employ a passive tracer to diagnose atmospheric transport and estimate $K_{zz}$ profiles in the kinetics runs. This tracer is solely advected by winds but does not influence radiative opacity or participate in chemical reactions. It is used to represent extremely long-lived chemical species with a uniform source from the deep atmosphere, whose distribution is primarily shaped by transport processes. We focus on the pressure levels where atmospheric transport dominates, so no source/sink terms are applied to the tracer field. Therefore, the evolution of tracer abundance in the upper atmosphere is governed by advection. 

To justify this approach, we compare $\tau_\mathrm{chem}$ of CO, CO$_2$, NH$_3$, and CH$_4$ from \citet{zahnle_methane_2014} with $\tau_\mathrm{mix}$ and $\tau_\mathrm{adv}$ derived from our synchronous simulation \citepalias[][fig.~1]{Liu_2025_PartI}. The results indicate that $\tau_\mathrm{chem} \gg \tau_\mathrm{mix}$ and $\tau_\mathrm{adv}$ at pressures smaller than $\sim$10~bar, suggesting that the transport-induced disequilibrium chemistry dominates these regions, and ignoring the source/sink due to chemical reactions at these regions is reasonable. This choice is supported by the fact that the quenched pressures for most of the dominant species range between 1 and 10 bar in 3D simulations, as shown below.  
The passive tracer is initialised to be globally uniform below 10~bar with a constant mass mixing ratio of $q_\mathrm{init} = 10^{-5}$. Above 10~bar, it follows a vertically decreasing profile with a power-law dependence:
\begin{equation}
    q_\mathrm{init} = 10^{-5} \times \left( \frac{P_\mathrm{bar}}{10~\mathrm{bar}} \right)^{1.5}.
\end{equation}
The estimated $K_{zz}$ profiles are expected to be insensitive to the exact value of $q_\mathrm{init}$ as long as the initial vertical gradient of the tracer is sufficiently steep, and the simulations run long enough to achieve a quasi-steady state \citep{komacek_vertical_2019}.

The kinetics runs are run for 5100 days, which is long enough to let the chemical abundances at the photosphere reach equilibrium, and the top of the atmosphere energy imbalances are within 5~W\,m$^{-2}$. The last 330~days (10 years for K2-18b) of the simulations with 10-day output frequency are averaged and used in the subsequent analysis.

\section{Results}\label{sec:results}
In this section, we present the results of 3D self-consistent hydrodynamics-radiative-chemistry simulations for the mini-Neptune K2-18b. We begin by summarising the thermal structure, atmospheric circulation, and passive tracer distribution, as previously described and analysed in \citetalias{Liu_2025_PartI} of this study. We then investigate the chemical structure in the simulations, followed by an estimation of the 1D $K_{zz}$ profile and a comparison between the results from 1D and 3D models. Finally, we present synthetic transmission spectra and compare them with the JWST observations reported by \citet{madhusudhan_carbon-bearing_2023} and \citet{Hu_K2-18b_2025}.

\subsection{Overview of the thermal structure, atmospheric circulation, and the passive tracer distribution}\label{subsec:overview}

We briefly revisit the thermal structure, atmospheric circulation patterns, and passive tracer distribution presented in \citetalias{Liu_2025_PartI}, as they provide the necessary context for the analysis that follows. For convenience, the thermal structure and wind patterns are provided in Appendix~\ref{sec:figures_PartI}, while a detailed description can be found in \citetalias{Liu_2025_PartI}.

The global mean air temperature profiles from the kinetics runs are consistent between simulations of different rotation periods (Fig.~\ref{fig:thermal_structure}a). 
They all exhibit an increase in air temperature with pressure below 5$\times$10$^{-4}$~bar, and a thermal inversion above 5$\times$10$^{-4}$~bar, due to the strong shortwave absorption of the abundant CH$_4$. 
The strong absorption of CH$_4$ and CO$_2$ in the near-infrared wavelengths induces a detached convective zone between 1 and 5~bar (Fig.~\ref{fig:thermal_structure}b). Vigorous convection can be triggered in this region, resulting in strong vertical mixing. In the heliocentric frame (with the substellar point fixed at 0$^\circ$ latitude and 0$^\circ$ longitude), a temperature difference between the evening (90$^\circ$ substellar longitude) and morning terminators (-90$^\circ$ substellar longitude) increases from a few Kelvin at 0.02~bar to 120~K at the model top \citepalias[][fig.~3e]{Liu_2025_PartI}. This difference results from the eastward wind dominating the upper atmosphere and is consistent across all the simulations.

Eastward wind dominates pressure levels above 0.1~bar, with the occurrence of equatorial superrotating jets, across all the simulations (Figs~\ref{fig:wind_pattern}a and b). Two high-latitude jets are present in the 6:1 SOR and 10:1 SOR simulations. At pressure levels above 0.01~bar, meridional wind is characterised by an equator-to-pole circulation on the day side and a reversed circulation at the night side (Figs~\ref{fig:wind_pattern}c and d). Upwelling dominates latitudes within $\pm$60$^\circ$ on the day side, and downwelling motions dominate the day-side high latitudes (Figs~\ref{fig:wind_pattern}e and f). These vertical wind patterns reverse on the night side.

Global mean vertical mixing strength and the global mean passive tracer abundance agree well among simulations with different rotation periods (Fig.~\ref{fig:mole_profile}f). However, the horizontal distribution of passive tracer abundances varies with rotation periods. Passive tracers are more abundant within $\pm$60$^\circ$ latitudes in the synchronous, 2:1 SOR, and 6:1 SOR simulations (Fig.~\ref{fig:mole_2D}a), positively correlated with the vertical wind. In the 10:1 SOR, passive tracers can still accumulate more at latitudes within $\pm$60$^\circ$, but are also abundant at high-latitude regions (>70$^\circ$, Fig.~\ref{fig:mole_2D}a), aligning with the zonal mean downwelling motion. This negative correlation between passive tracer abundance and the vertical wind at high latitudes in the 10:1 SOR simulation is caused by the strong vertical transient eddies that can transport passive tracer upward \citepalias[][fig.~10a]{Liu_2025_PartI}. Longitudinally, the strong eastward wind transports passive tracers eastward from the substellar point, where the vertical mixing is the strongest, leading to $\sim$20\% more abundant passive tracers at the evening terminator than the morning terminator (Fig.~\ref{fig:mole_2D}a).

\subsection{Chemical structure}\label{subsec:chemical}

\begin{figure*}
\includegraphics[width=2\columnwidth]{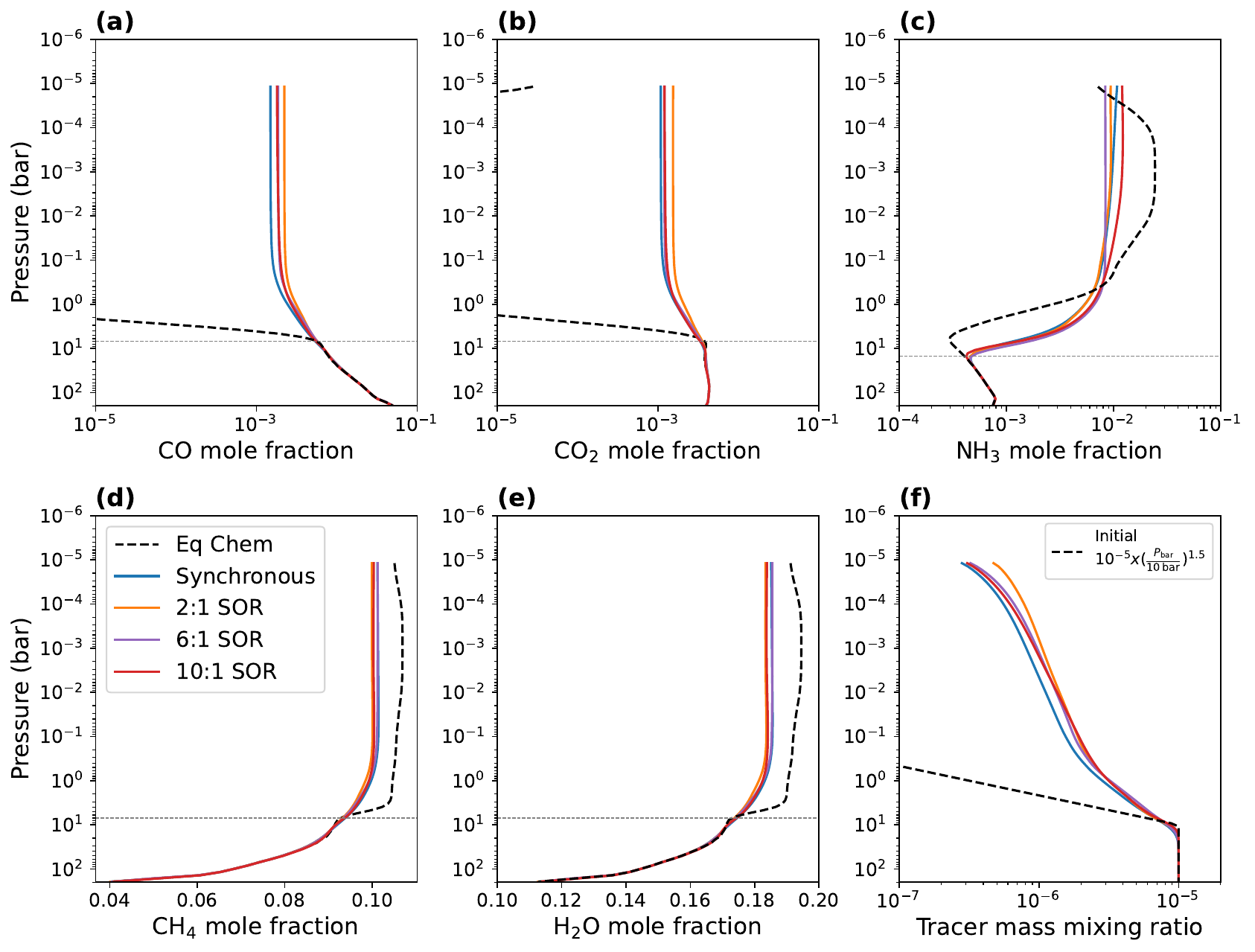}
\caption{
Global-mean vertical profiles of CO (a), CO$_2$ (b), NH$_3$ (c), CH$_4$ (d), and H$_2$O (e) mole fractions and passive tracer mass mixing ratio (f).
Coloured lines show results from simulations assuming K2-18b is a synchronous rotator (blue), or in 2:1 SOR (orange), 6:1 SOR (purple), and 10:1 SOR (red). 
Black dashed lines depict the chemical equilibrium abundances calculated via Gibbs free energy minimisation using the temperature field from the kinetics run; because the global-mean temperature profiles are consistent across different rotation rates, only results from the synchronous simulation are shown here. 
Horizontal grey dotted lines indicate the quench levels for each species. The quench levels for CO, CO$_2$, NH$_3$, CH$_4$, and H$_2$O are 7, 7, 15, 7, and 7 bar, respectively. The dashed black line in panel (f) indicates the initial passive tracer mass mixing ratio used in all simulations.
}
\label{fig:mole_profile}
\end{figure*}

In this section, we examine the chemical structure resulting from our simulations. Fig.~\ref{fig:mole_profile} presents the global-mean vertical profiles of the dominant chemical species that contribute to the transmission spectra. The coloured lines represent the abundances under transport-induced disequilibrium chemistry. The black dashed lines show the chemical equilibrium abundances calculated via Gibbs free energy minimisation based on the temperature field from the synchronous kinetics run. Unless otherwise noted, we refer to the black dashed line results as the `chemical equilibrium' case, as they are based on the same temperature profiles as the chemical kinetics–transport simulations. We note that the global-mean temperature profiles remain nearly unchanged after enabling chemical kinetics–transport, so the final equilibrium abundances closely agree with the initial field (not shown).

Since the global mean air temperature profiles remain consistent across different cases (mentioned above, shown in Fig.~\ref{fig:thermal_structure}a, the mole fraction of chemical species remains invariant under local chemical equilibrium, and only results from the synchronous simulation are shown in Fig.~\ref{fig:mole_profile}. Overall, under chemical equilibrium, CO and CO$_2$ abundances exhibit the largest vertical gradient, varying over ten orders of magnitude; NH$_3$ mole fraction can vary by one order of magnitude; and the changes in CH$_4$ and H$_2$O abundances are less significant, varying by less than one order of magnitude over the 200 bar pressure range.

As shown by the dashed lines in Figs~\ref{fig:mole_profile}a, b, d, and e, under the assumption of chemical equilibrium, CO and CO$_2$ are more abundant in regions of high temperature, while CH$_4$ and H$_2$O are more abundant under low-pressure and low-temperature conditions. CO and CO$_2$ abundances exhibit steep vertical gradients between approximately 10 and 5$\times$10$^{-4}$~bar, decreasing from around $10^{-2}$ at 10~bar to about $10^{-24}$ for CO and $10^{-19}$ for CO$_2$ at 5$\times$10$^{-4}$~bar (these values are too small to be visible in Figs~\ref{fig:mole_profile}a and b). Above this level, their abundances increase with altitude, reaching approximately $10^{-7}$ for CO and $10^{-5}$ for CO$_2$ at the top of the atmosphere. The reversal in the vertical abundance trends is caused by the temperature inversion that occurs above 5$\times$10$^{-4}$~bar (Fig.~\ref{fig:thermal_structure}a). The equilibrium abundances of CH$_4$ and H$_2$O increase with altitude between approximately 10 and 5$\times$10$^{-4}$~bar, but slightly decrease above 5$\times$10$^{-4}$~bar due to the temperature inversion at these high altitudes.

The above trends with temperature can be understood by analysing the interconversion of \ce{CH4 <=> CO} and \ce{CO <=> CO2}. In the H$_2$-dominated atmosphere, the interconversion between CH$_4$ and CO is governed by the net reaction: \ce{CH4 + H2O <=> CO + H2}  \citep[e.g.][]{Burrows_1999,yung1999photochemistry,Moses_2011,Madhu_2012_CtoO}. At high temperatures ($\gtrsim$1000~K, corresponding to pressures above $\sim$10 bar in our simulations), the conversion of CH$_4$ to CO is favoured, while at low temperatures ($\lesssim$1000~K, corresponding to pressures below $\sim$10 bar in our simulations), the reverse reaction is favoured. Consequently, CH$_4$ mole fraction decreases with increasing temperature, while CO increases with increasing temperature \citep[e.g.][]{Madhu_2012_CtoO,Mose_2013} and is depleted in high altitudes. CO$_2$ is mainly converted from CO and is in pseudo-equilibrium with CO \citep[e.g.][]{yung1999photochemistry,Moses_2011,Tsai_2018}, so that as CO depletes in low temperatures, CO$_2$ also depletes. As CO and CO$_2$ deplete in low temperatures, H$_2$O becomes the dominant oxygen-bearing species, and its mole fraction enhances with decreasing temperature. 

Under chemical equilibrium, NH$_3$ is more abundant under low-pressure and low-temperature conditions (Fig.~\ref{fig:mole_profile}c). Its abundance increases from $<$10$^{-3}$ in the deep atmosphere (P$>$10~bar) to $>$10$^{-2}$ in the cooler upper atmosphere (P$<$0.1~bar). Between approximately 10 bar and 5$\times$10$^{-4}$~bar, its equilibrium abundances increase with altitude, but decrease above 5$\times$10$^{-4}$~bar due to the temperature inversion at these high altitudes. This trend is because both the conversion between N-bearing species: \ce{N2 + 3H2 <=> 2NH3} and \ce{HCN + 3H2 <=> NH3 + CH4} are favoured to produce NH$_3$ at lower temperatures \citep[e.g.][]{Moses_2011}.

Under transport-induced disequilibrium chemistry, the global mean mole fractions of chemical species exhibit some variations across different rotation periods (Fig.~\ref{fig:mole_profile}). However, these changes remain minor, for example, the CO$_2$ mole fraction is 1.1$\times$10$^{-3}$, 1.5$\times$10$^{-3}$, 1.2$\times$10$^{-3}$, and 1.2$\times$10$^{-3}$ for synchronous, 2:1 SOR, 6:1 SOR, and 10:1 SOR simulations, respectively. This suggests that rotation periods do not significantly impact vertical transport, which agrees with the results implied by the passive tracer as discussed in \citetalias{Liu_2025_PartI} (Fig.~\ref{fig:mole_profile}f). 

The limited influence of rotation period on vertical transport is primarily attributed to the similarity of the mean circulation patterns across all simulations -- in particular, the day-side upwelling structures are consistently present regardless of the rotation period (Figs~\ref{fig:wind_pattern}e and f). While eddies can also transport tracers vertically, their contribution to the globally averaged vertical flux remains secondary compared to the mean circulation. The vertical transport of the passive tracer is governed by 

\begin{align}
\frac{\partial{([\overline{\rho}]}[\overline{q}])}{
\partial{t}}\sim  & -\frac{1}{r^2}\left[\frac{\partial([\overline{\rho w}][\overline{q}]r^2)}{\partial r}
+\frac{\partial([\overline{(\rho w)'q'}]r^2)}{\partial r}\right. \nonumber\\ & \left.+\frac{\partial([\overline{\rho w}^*\overline{q}^*]r^2)}{\partial r}\right]\,\,,
\end{align}

\noindent where the three terms on the right-hand side represent vertical transport by the mean flow, transient eddies, and stationary eddies, respectively  \citepalias[][equation~9]{Liu_2025_PartI}. 
Comparing the global mean absolute values of each term, we find that the mean-flow transport term dominates over the stationary eddy term by approximately 4, 7, 6, and 4 orders of magnitude, and over the transient eddy term by approximately 4, 4, 4, and 3 orders of magnitude, for the synchronous, 2:1 SOR, 6:1 SOR, and 10:1 SOR simulations, respectively.

The molecular abundances of CO and CO$_2$ begin to significantly deviate from chemical equilibrium under transport-induced chemistry at approximately 7 bar, where quenching happens (Figs~\ref{fig:mole_profile}a and b). 
Above this level, the dynamical timescale becomes shorter than their chemical timescales, and atmospheric transport dominates the distribution of chemical species, leading to substantial changes in their abundances. As a result, CO and CO$_2$ mole fractions increase to 10$^{-3}$ above 0.1~bar due to vertical transport from deeper layers. As noted above, the chemical kinetics–transport simulations are initialised using chemical equilibrium abundances. This setup makes CO and CO$_2$ behave similarly to the deep-sourced passive tracer used in this study under atmospheric transport, albeit with non-uniform sources (discussed below).

For NH$_3$, the quench level is around 15~bar (Fig. \ref{fig:mole_profile}c). Above this level, chemical reactions are too slow to maintain local equilibrium, atmospheric transport effectively homogenises the NH$_3$ abundance, smoothing its vertical gradient. As we initialise the simulations from chemical equilibrium conditions where NH$_3$ is more abundant in the upper atmosphere ($\sim$1 bar) than in the deeper region (1–15 bar). Atmospheric transport therefore drives a net downward flux of NH$_3$. This enhances NH$_3$ in the deeper, low-abundance region (1--15~bar) while depleting it above $\sim$1~bar. Compared to chemical equilibrium, the NH$_3$ mole fraction in the photosphere decreases from $\sim$0.025 to $\sim$0.010 under transport-induced disequilibrium chemistry.

The quench levels for CH$_4$ and H$_2$O are around 7~bar. Above the quench level ($\sim$7 bar), the thermochemical timescales of CH$_4$ and H$_2$O are much larger than the dynamical timescale, resulting in vertical mixing dominating their distribution. Since both species are more abundant above the quench level under initial thermochemical equilibrium than at the quench level, this prompts atmospheric circulation to drive a net downward transport and depletes CH$_4$ and H$_2$O above the quench level (Figs~\ref{fig:mole_profile}d and e). However, the effect is modest compared to NH$_3$, owing to the shallower vertical gradients of CH$_4$ and H$_2$O in chemical equilibrium. The mole fraction of CH$_4$ decreases from 0.11 to 0.10, and that of H$_2$O decreases from 0.19 to 0.18 in the photosphere under transport-induced disequilibrium chemistry compared to chemical equilibrium.

\begin{figure*}
\includegraphics[width=2\columnwidth]{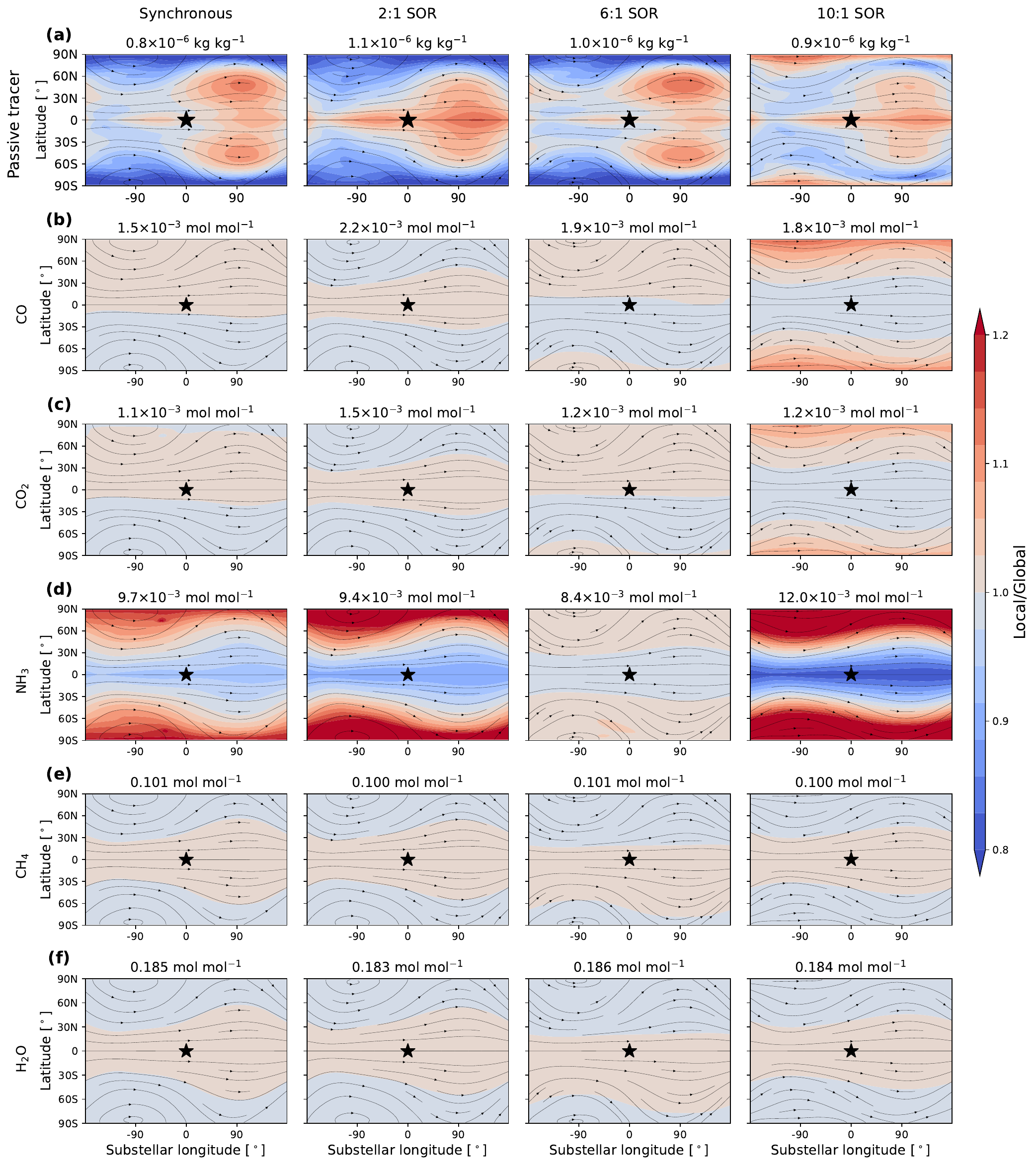}
\caption{Horizontal distribution of passive tracer (a), CO (b), CO$_2$ (c), NH$_3$ (d), CH$_4$ (e), and H$_2$O (f) at 0.001 bar. The results are transformed into the heliocentric frame, keeping the substellar point at 0$^\circ$ longitude and 0$^\circ$ latitude. The black star-shaped markers indicate the location of the substellar point. The streamlines indicate the direction of the flow.
To highlight the horizontal gradient, the contours indicate the local mole fraction divided by the global mean at this pressure level. The corresponding average values are displayed above each panel. 
Columns from left to right show results from synchronous, 2:1 SOR, 6:1 SOR, and 10:1 SOR simulations, respectively.}
\label{fig:mole_2D}
\end{figure*}

\begin{figure*}
\centering
\includegraphics[width=1\columnwidth]{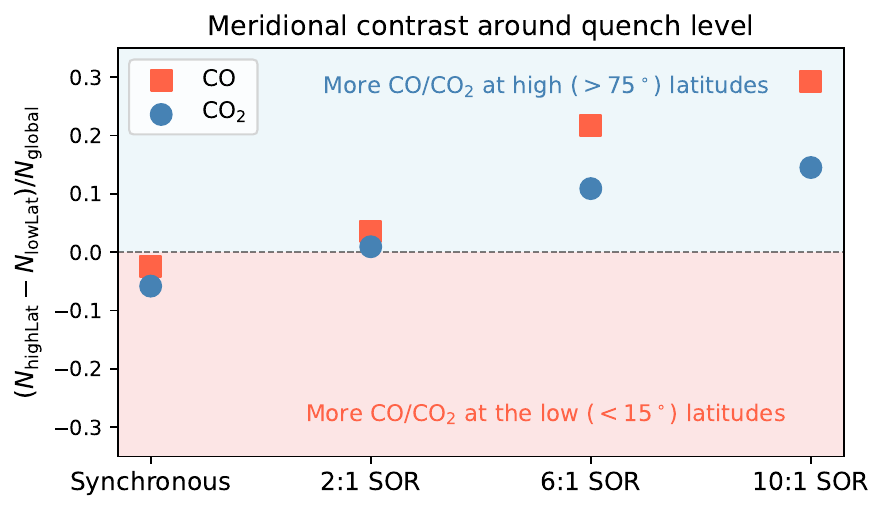}
\caption{Column-integrated CO and CO$_2$ mole fraction contrast between high- ($>75^\circ$) and low- ($<15^\circ$) latitude regions around quench level. The vertical axis shows the difference between the high- and low-latitude column abundances normalised by the global mean. Orange squares and blue circles denote CO and CO$_2$, respectively. Positive values indicate that CO and CO$_2$ are more abundant at high-latitude regions.}
\label{fig:mole_quench}
\end{figure*}

The horizontal distribution of chemical species at 0.001~bar is shown in Fig.~\ref{fig:mole_2D}. We present the horizontal chemical structures in a heliocentric frame, where the substellar point remains fixed at 0$^\circ$ longitude and 0$^\circ$ latitude. To highlight the horizontal gradient, the contours indicate the local mole fraction divided by the global mean at this pressure level. 

Unlike the passive tracer (Fig.~\ref{fig:mole_2D}a), which exhibits pronounced longitudinal variations, the chemical species show largely uniform abundances along longitudes in the upper atmosphere (Figs~\ref{fig:mole_2D}b--f). This difference arises from the distinct initial conditions of the two types of tracers.

The passive tracer has an idealised uniform horizontal distribution and a steep, monotonic vertical gradient initially, with high abundances in the deep atmosphere and depletion in the upper atmosphere. Consequently, vertical motions driven by atmospheric circulation efficiently transport tracer-rich air upward and tracer-poor air downward, generating strong longitudinal contrasts.
In contrast, chemical species cannot trace vertical mixing as directly as the passive tracer for two reasons: (i) Chemical species exhibit relatively weak or non-monotonic vertical gradients in the upper atmosphere under initial thermochemical equilibrium, which limit the horizontal contrast generated by the vertical displacement. For example, CO and CO$_2$ decrease with altitude from 10 bar to 5$\times$10$^{-4}$ bar, but increase above 5$\times$10$^{-4}$ bar due to the temperature inversion. At 0.001 bar, mixing of CO and CO$_2$ with the more CO/CO$_2$-abundant upper atmosphere can modify the chemical contrast established by vertical transport from deeper layers. Meanwhile, NH$_3$, CH$_4$, and H$_2$O vary by only about one order of magnitude vertically under initial equilibrium conditions, limiting the horizontal contrasts generated by vertical displacements. (ii) Under initial thermochemical equilibrium, the non-uniform temperature field produces longitudinal inhomogeneities in chemical species distributions, which can partially offset the contrasts generated by vertical transport. For example, at 0.001 bar, CO and CO$_2$ reach their maximum abundances in the hottest region east of the substellar point (70$^\circ$ longitude, figure not shown), which is offset by 20$^\circ$ from the peak passive tracer abundance region (90$^\circ$ longitude, Fig.~\ref{fig:mole_2D}a). This spatial offset reduces the chemical contrast that would be induced by atmospheric transport alone.
Together, these factors result in more uniform distributions of chemical species in the upper atmosphere compared to the passive tracer.

The coloured contours in Figs~\ref{fig:mole_2D}b--f broadly follow the wind streamlines and exhibit a fish-like morphology, with a prominent `head' near the evening terminator (90$^\circ$) and a narrowing `tail' extending toward the morning terminator (-90$^\circ$). This pattern further suggests that horizontal winds predominantly shape the longitudinal distribution of chemical species and smooth their contrast. Nevertheless, meridional gradients in chemical abundances remain evident.

Overall, NH$_3$ exhibits the strongest meridional gradient among the five chemical species, with $\sim$50\% changes in synchronous rotator and 2:1 SOR simulations, $\sim$10\% in the 6:1 SOR simulation, and up to $\sim$100\% in the 10:1 SOR simulation (Fig.~\ref{fig:mole_2D}d), discussed below. The meridional gradients of CO and CO$_2$ remain modest, with abundance variations within 2\% for the synchronous rotator and 2:1 SOR simulations, within 10\% for the 6:1 SOR case, and 20\% for the 10:1 SOR case (Figs~\ref{fig:mole_2D}b and c). The abundances of CH$_4$ and H$_2$O remain relatively homogeneous across the horizontal domain, exhibiting variations of less than 2\% in all the simulations (Figs~\ref{fig:mole_2D}e and f).

CO and CO$_2$ are more abundant at latitudes within $\pm50^\circ$ in the 2:1 SOR simulation, whereas in the 6:1 SOR and 10:1 SOR simulations, their concentrations peak at higher latitudes (Figs~\ref{fig:mole_2D}b and c). In the synchronous simulation, CO and CO$_2$ exhibit slightly higher abundances in the northern hemisphere, but this hemispheric asymmetry is minor ($<$1\%). Although the initial setup of CO and CO$_2$ is similar to that of the passive tracer as mentioned above, their distributions are more complex. Unlike the passive tracer, which is initialised with a globally uniform abundance, CO and CO$_2$ exhibit some spatial variations in abundance at the quench level (Fig.~\ref{fig:mole_quench}), making their meridional distribution affected by both the vertical mixing and their distribution at the quench level.

In the 2:1 SOR simulation, CO and CO$_2$ are slightly more abundant at low latitudes due to the prevalence of upwelling motions in these regions, which aligns with the positive correlation between passive tracer abundance and vertical velocity as discussed in \citetalias{Liu_2025_PartI}. In contrast, in the 6:1 SOR simulation, the enhanced abundances of CO and CO$_2$ at high latitudes are primarily driven by their elevated abundances around the quench level in those regions (Fig.~\ref{fig:mole_quench}). Although the passive tracer abundance correlates positively with vertical wind in the 6:1 SOR simulation, the distribution of species at the quench level dominates in shaping the meridional abundance patterns. In the 2:1 SOR simulation, although CO and CO$_2$ are also more abundant at high latitudes at the quench levels, the meridional contrasts are small, so the vertical wind pattern still dominates the meridional abundance patterns  (Fig.~\ref{fig:mole_quench}). In these two simulations, the effects of vertical transport and the quench-level distribution act in opposite directions, largely cancelling each other and resulting in weak meridional contrasts.

In the 10:1 SOR simulation, CO and CO$_2$ are also more abundant at high latitudes, driven by two factors: first, vertical transient eddies in the high-latitude regions enhance upward transport; second, CO and CO$_2$ are intrinsically more abundant at high latitudes around the quench level (Fig.~\ref{fig:mole_quench}). These factors reinforce each other, leading to a larger meridional contrast compared to the 2:1 SOR and 6:1 SOR simulations.

The higher quenched abundances of CO and CO$_2$ at high latitudes in the 2:1 SOR, 6:1 SOR, and 10:1 SOR simulations are due to the warmer temperatures relative to the lower latitudes (Figs~\ref{fig:tair_7bar}b--d). The warmer polar regions at this layer are likely due to eddy heat flux divergence at this layer (Figs~\ref{fig:eddy_heat}b--d). However, the vertical resolution in the 5--10~bar region is significantly coarser than in the upper atmosphere: this pressure range contains only $\sim$5 model levels with an average spacing of $\sim$1~bar, compared to $\sim$10 levels with an average spacing of $\sim$0.01~bar in the 0.001--0.1~bar range -- a factor of $\sim$100 difference in resolution. As a result, the temperature contrasts identified in this region could also arise from interpolation errors or unresolved dynamical processes rather than a physically robust signal. A more detailed investigation with higher vertical resolution is required to accurately determine the underlying cause in future work.

The minor hemispheric asymmetry ($<$1\%) in the CO and CO$_2$ distributions in synchronous simulation, despite the model being otherwise symmetric with no imposed obliquity or asymmetric forcing, may be attributed to numerical diffusion in the dynamical core rather than any physical mechanism. The almost uniform distribution of CO and CO$_2$ suggests that vertical and horizontal mixing can effectively smooth the contrast. This may be partly due to the initially non-monotonic vertical profiles of CO and CO$_2$, which lack a consistent gradient direction with altitude; vertical transport driven by atmospheric dynamics, therefore, may tend not to systematically enrich or deplete these species in a particular region, resulting in relatively small horizontal contrasts.

NH$_3$ consistently shows higher abundances at high latitudes ($>$50$^\circ$) across all simulations. This pattern arises primarily from two factors. First, NH$_3$ is more abundant at high latitudes initially under chemical equilibrium, as it is more thermodynamically stable in cooler environments. Since the chemical timescale of NH$_3$ is long compared to the dynamical timescale above the quench level ($\sim$15~bar), this chemical equilibrium latitudinal contrast largely serves as the baseline spatial pattern upon which dynamical transport acts. This latitudinal contrast in the initial chemical equilibrium is $10-20\%$ above quench level in the synchronous, 2:1 SOR, and 6:1 SOR simulations, and is $\sim60\%$ in the 10:1 SOR simulation (Fig.~\ref{fig:NH3_change}). Second, because the initial vertical NH$_3$ profile increases with altitude below 5$\times$10$^{-4}$~bar, large-scale overturning circulation further amplifies the meridional contrast: downwelling motions at high latitudes transport NH$_3$-rich upper atmospheric air downward, while upwelling motions at low latitudes bring NH$_3$-poor deep atmospheric air upward. This is evident from the final polar–equatorial contrast, which increases under disequilibrium chemistry across all rotations (Fig.~\ref{fig:NH3_change}). The combination of these two factors, namely a large initial latitudinal contrast in chemical equilibrium and a dynamically amplified vertical redistribution, gives NH$_3$ the most pronounced meridional contrast among all species considered.

CH$_4$ and H$_2$O are slightly more abundant at low latitudes, likely due to the combined effects of initial meridional abundance gradients and the large-scale circulation. However, the resulting latitudinal contrast is weak, with variations of less than 2\% in all simulations. Compared to NH$_3$, CH$_4$ and H$_2$O do not exhibit large meridional contrast due to two reasons. First, their chemical equilibrium distributions show little latitudinal variation above the quench level ($<5\%$), providing a much smaller initial spatial contrast. Second, their vertical abundance gradients are also significantly smaller than that of NH$_3$, meaning that vertical transport by the mean circulation produces a comparatively modest meridional redistribution.

In summary, the meridional distribution of the chemical species in the upper atmosphere is the combined result of chemical calculation and atmospheric dynamics. Chemical processes set the initial meridional contrasts at the quench levels and determine the sources of individual species, while atmospheric motions act to smooth these contrasts vertically.
Despite this smoothing, persistent horizontal variations remain, highlighting the need for fully coupled 3D chemical–dynamic simulations to accurately capture the 3D distribution of atmospheric composition.

\subsection{\texorpdfstring{$K_{zz}$}{Kzz} estimation and comparison to 1D models}\label{subsec:Kzz}

In this section, we estimate 1D $K_{zz}$ profiles that characterise the strength of globally-averaged vertical mixing. We then compare the 1D model results, using these $K_{zz}$ profiles, with the full 3D model outputs. This approach allows us to infer the effective vertical diffusivity, which, in the context of a 1D model, would produce the same net vertical transport as that driven by dynamical mixing processes in the 3D model, such as large-scale circulation and small-scale eddies.

\begin{figure*}
\centering
\includegraphics[width=1.9\columnwidth]{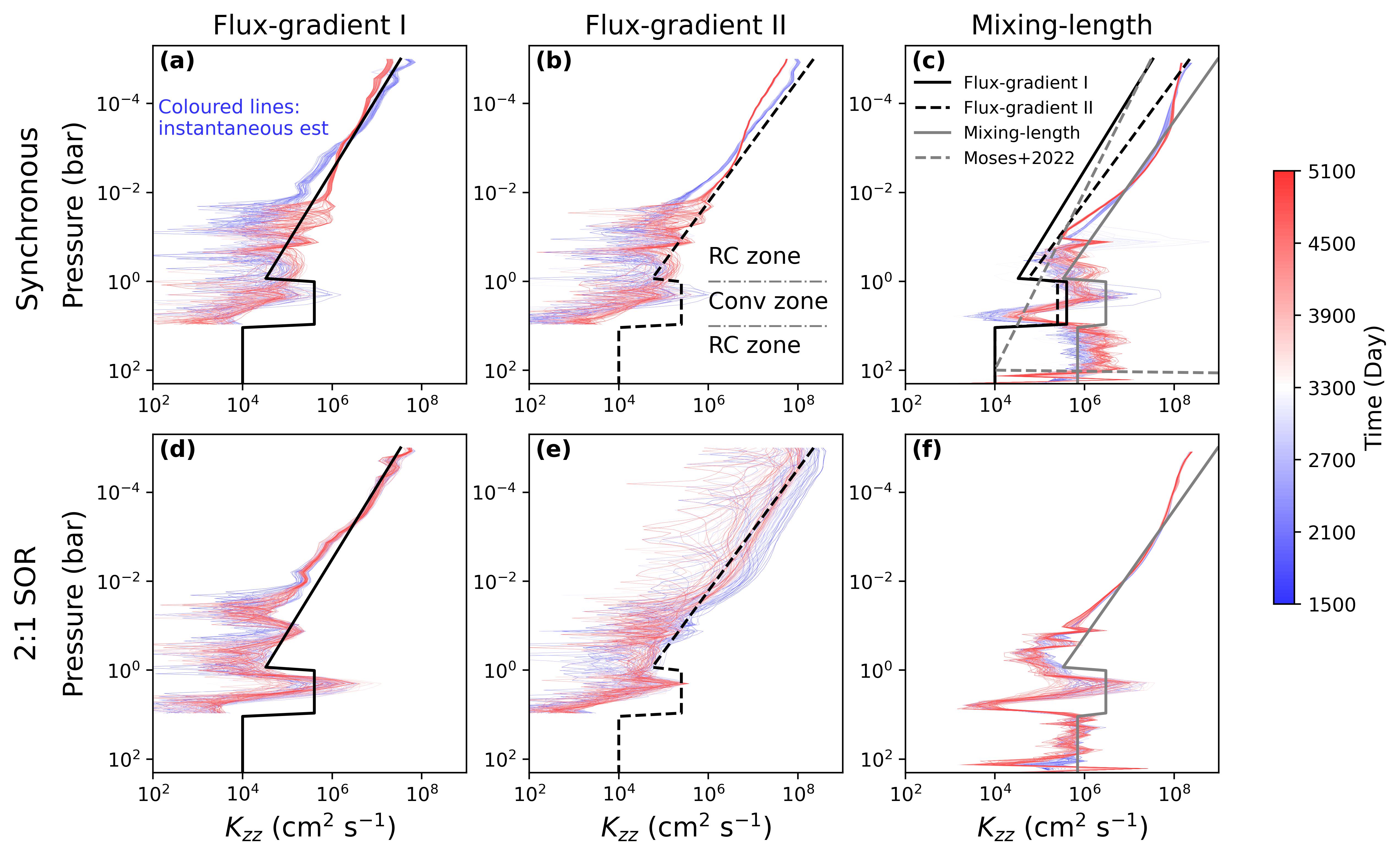}
\caption{$K_{zz}$ derived with the flux-gradient relationship I (a) and II (b) and the mixing-length theory (c). Panels (a)--(c) show results from the synchronous simulation, while panels (d)--(f) correspond to the 2:1 SOR simulation. Coloured lines represent the $K_{zz}$ profiles calculated every thirty model days from the instantaneous model output}, where bluer lines represent earlier stages of the simulation and redder lines indicate later stages. The results for the last 3600 days are shown after the simulations have spun up.
The thick solid (dashed) black and solid grey lines represent a roughly `average' parametrisation for the $K_{zz}$ profiles derived from the flux-gradient relationship I (II) and the mixing length theory. The three average $K_{zz}$ profiles estimated from different methods are all plotted in panel (c) for comparison. The thin grey dashed-dotted lines indicate the boundary between the radiative-convective (RC) zones and the convective (Conv) zone in panel (b).
Above 1~bar, the average $K_{zz}$ profiles are  
$3\times10^4\cdot(1~\text{bar}/P_\text{bar})^{0.61}$~cm$^2$~s$^{-1}$, $5\times10^{4}\cdot(1~\text{bar}/P_\text{bar})^{0.7}$~cm$^2$~s$^{-1}$, and $3\times10^{5}\cdot( 1~\text{bar}/P_\text{bar})^{0.7}$~cm$^2$~s$^{-1}$ for flux-gradient relationship I (solid black), flux-gradient relationship II (dashed black), and the mixing length theory (solid grey), respectively. Between 1 and 10~bar, the average $K_{zz}$ profiles are set to be constants: 4$\times$10$^{5}$~cm$^2$~s$^{-1}$,  4$\times$10$^{5}$~cm$^2$~s$^{-1}$, and 3$\times$10$^{6}$~cm$^2$~s$^{-1}$ for the three methods. Note that the detached convective zone is set to be more extended vertically, from 1 to 10~bar in the average $K_{zz}$ profile. The grey dashed line in 
panel (c) is the estimation in \citet[][equation 1]{moses2022chemical}: $K_{zz} = 9.77 \times 10^4 \cdot (1~\mathrm{bar}/P_\text{bar})^{0.5}$~cm$^2$\,s$^{-1}$.
\label{fig:Kzz_profile}
\end{figure*}

\begin{figure*}
\centering
\includegraphics[width=1.7\columnwidth]{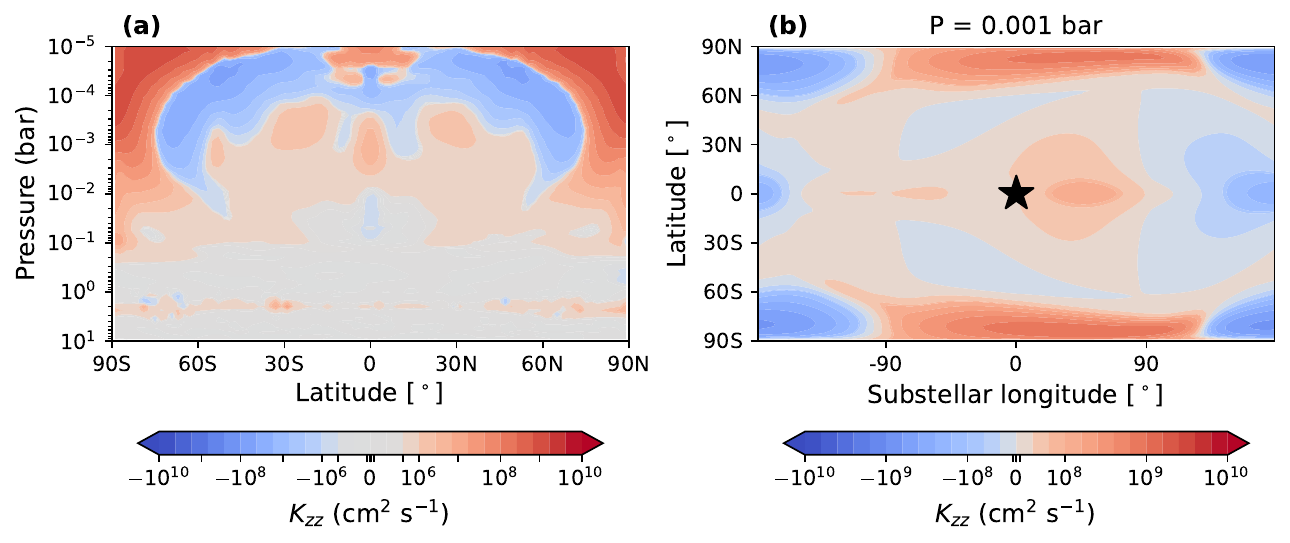}
\caption{The spatial distribution of $K_{zz}$ estimated using the flux-gradient relationship II (equation \ref{eq:flux-gradient II}). (a) The latitude versus pressure distribution of zonal mean $K_{zz}$. (b) The horizontal distribution of $K_{zz}$ at 0.001 bar. The black star-shaped marker indicates the location of the substellar point. $K_{zz}$ shown in this figure is derived from the synchronous simulation, averaged over the last 3600 days after it spins up, with an output interval of 10 days.}
\label{fig:Kzz_2D}
\end{figure*}

\begin{figure*}
\centering
\includegraphics[width=2\columnwidth]{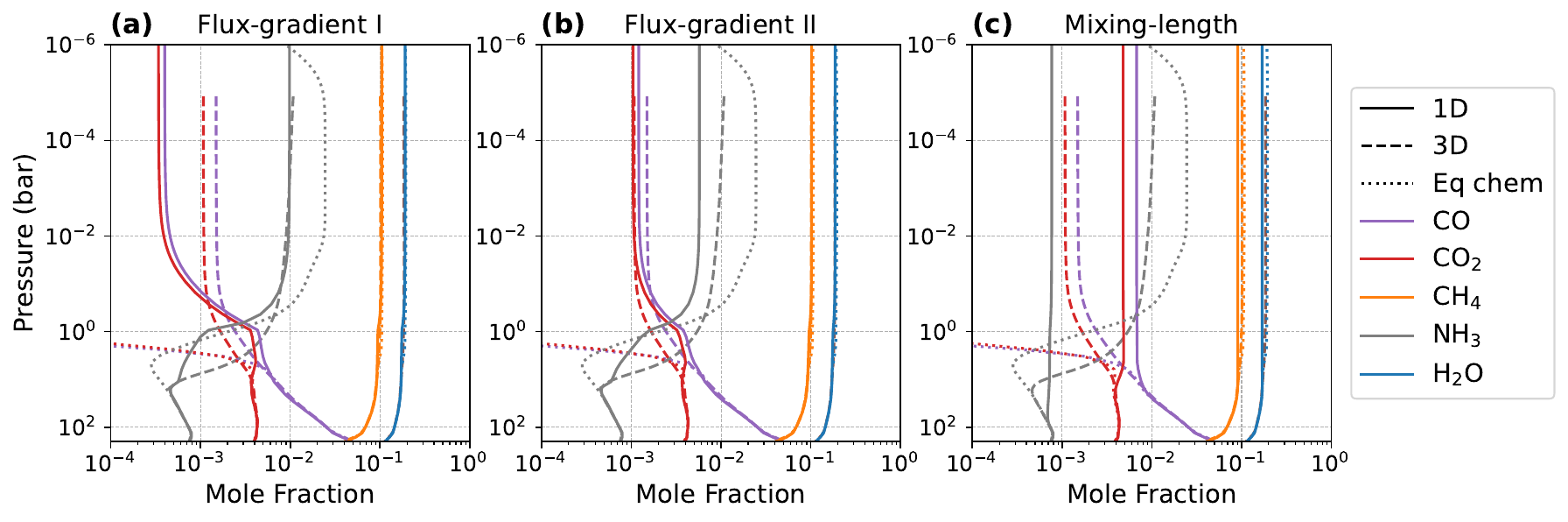}
\caption{The comparison between the chemical species abundances from 1D \texttt{ATMO} simulations (solid) employing different $K_{zz}$ profiles and the global-mean results in the synchronous 3D simulation (dashed). The three panels are results with the average $K_{zz}$ profile derived from the flux-gradient relationship I (a), flux-gradient relationship II (b), and the mixing length theory (c) as shown in Fig. \ref{fig:Kzz_profile}.}
\label{fig:3D_vs_1D}
\end{figure*}

We estimate the global mean $K_{zz}$ profiles from the GCM simulations using two different methods: the flux-gradient relationship from the passive tracer and the mixing-length theory \citep[e.g.][]{mccormac_vertical_1975,plumb1987zonally,zhang_global-mean_2018}, discussed below. The global-mean $K_{zz}$ values estimated using the two methods are presented in Fig.~\ref{fig:Kzz_profile}, with estimates from the synchronous rotator and 2:1 SOR simulations shown as examples, as the other simulations yield similar profiles. 

For the flux-gradient relationship, we use two equations that have been documented and used in previous studies. The first is based on the tracer mass flux \citep[e.g.][]{mccormac_vertical_1975,Chamberlain_1987,parmentier_3d_2013,komacek_vertical_2019}:
\begin{equation}
    K_{zz} = -\frac{\langle\rho q w \rangle}{\langle\rho \frac{\partial q}{\partial r}\rangle},
\label{eq:flux-gradient I}
\end{equation}
where $\rho$ and $r$ are the air density and radial coordinate, and the brackets denote the horizontal average along isobars. We refer to equation \ref{eq:flux-gradient I} as `flux-gradient relationship I'. 

The second equation is based on the eddy tracer mixing ratio flux, \citep[e.g.][]{plumb1987zonally,zhang_global-mean_2018,zhang_2018,Steinrueck_2021}:
\begin{equation}
    K_{zz} = -\frac{\langle w'q' \rangle}{ \frac{\partial \langle q\rangle}{\partial r}},
\label{eq:flux-gradient II}
\end{equation}
where $q'$ and $w'$ are the perturbations of the passive tracer mass mixing ratio and vertical velocity from the horizontal average. 
We refer to equation \ref{eq:flux-gradient II} as `flux-gradient relationship II'. 

The flux-gradient relationship links the tracer flux to the vertical gradient of the mean value, inferring an effective diffusive vertical mixing rate from an inherently non-diffusive three-dimensional atmosphere. 
The theoretical background of this approach is that we assume that the tracer is mixed upward by the large-scale overturning circulation. If the tracer distribution is initially horizontally uniform but exhibits a vertical gradient in its global-mean abundance, vertical transport will naturally generate tracer perturbations on isobaric surfaces that are correlated with the vertical velocity field. Regions of upward motion become enriched in tracers, while regions of downward motion become depleted.
Then the upward region can mix more tracer upward, and the downward region mixes less tracer downward. In a global-mean perspective, the atmospheric transport behaves in a diffusive-like process. It is most valid when the atmosphere is stratified and behaves diffusively, which requires horizontal tracer anomalies to be small \citep{zhang_global-mean_2018,zhang_2018}. In our simulations, however, the atmospheric transport deviates from ideal diffusive behaviour. The material surfaces of extremely long-lived tracers become highly distorted, causing the conventional eddy-diffusion framework to break down. This breakdown is evidenced by the occasional local negative $K_{zz}$ values (discussed below), which lead to negative global-mean $K_{zz}$ in some cases. In light of this, we present the absolute values of $K_{zz}$ in the subsequent analysis.

The estimation based on the mixing-length theory is 
\begin{equation}
    K_{zz}  \approx \overline{w}^2 \tau_{\text{adv}}.
\label{eq:mixing_length_theory}
\end{equation}
Here, $\overline{w}$ is the global root-mean-square of vertical velocity, and $\tau_{\text{adv}}$ is the horizontal timescale, estimated as $R_\mathrm{p}/\overline{u}$, where $R_\mathrm{p}$ is the planetary radius and $\overline{u}$ is the root-mean-square of zonal wind speed. This equation is simplified by ignoring the extremely long $\tau_{\text{chem}}$ as shown in \citet[][equation 13]{zhang_global-mean_2018} and \citet[][equation 30]{komacek_vertical_2019}. It holds when the scale of horizontal wind to horizontal length (the planetary radius) is comparable to the scale of vertical wind to the scale height ($\mathcal{U}/R_\mathrm{p} \sim \mathcal{W}/H$).

The $K_{zz}$ profiles (coloured lines) are computed every thirty days based on instantaneous model output. 
Comparing the three rows, the $K_{zz}$ values estimated using the mixing-length theory (Figs~\ref{fig:Kzz_profile}c and f) are generally one to two orders of magnitude larger than those derived from the flux-gradient relationship (Figs~\ref{fig:Kzz_profile}a, b, d, and e). This finding is consistent with \citet{parmentier_3d_2013}, where the $K_{zz}$ derived from mixing-length theory was found to exceed the tracer-derived estimates by about two orders of magnitude. 
The $K_{zz}$ profiles obtained from flux-gradient relationships I and II are in broad agreement, with the values from relationship II being slightly larger than those from relationship I above $\sim$0.001~bar.

In addition, the $K_{zz}$ profiles estimated at different times during the simulations exhibit large variations, suggesting that the strength of vertical transport can have significant temporal variability (coloured lines in Fig.~\ref{fig:Kzz_profile}). This behaviour is expected, as the estimates are derived from instantaneous model outputs rather than time-averaged fields.  Moreover, $K_{zz}$ also exhibits large spatial variations, leading to dips and spikes in the profiles, particularly in the estimation using the flux-gradient relationship (Figs~\ref{fig:Kzz_profile}a, b, d, and e). 

As shown in Fig.~\ref{fig:Kzz_2D}, the absolute value of $K_{zz}$ has large spatial variation; this variation can reach more than 3 orders of magnitude at a given level (Fig.~\ref{fig:Kzz_2D}b). For instance, at 0.001 bar, $K_{zz}$ reaches $\sim$10$^{10}$~cm$^2$~s$^{-1}$ near 90$^\circ$N at a longitude of 90$^\circ$, while dropping below 10$^{7}$~cm$^2$~s$^{-1}$ to the west of the substellar point at the same pressure level (Fig.~\ref{fig:Kzz_2D}b).
The large $K_{zz}$ values at polar regions are caused by the strong downwelling at the dayside polar regions (Fig.~\ref{fig:wind_pattern}e) and the reduced passive tracer abundance in the polar regions (Fig.~\ref{fig:mole_2D}a). The negative $K_{zz}$ values (blue regions) arise from distortions of the passive-tracer material surfaces due to horizontal transport. For example, at the evening terminator (90$^\circ$ longitude, Fig.~\ref{fig:Kzz_2D}b), $K_{zz}$ becomes negative because the tracer concentration is enhanced ($q'>0$) as eastward winds transport tracer-rich air from the sub-stellar region (Fig.~\ref{fig:mole_2D}a), where strong heating drives upwelling (Fig.~\ref{fig:wind_pattern}e). Downstream of this, the flow becomes downward, giving $w'<0$. These combined produce negative $w'q'$, and therefore negative local $K_{zz}$.

The $K_{zz}$ profiles estimated from different methods are similar in shape. They exhibit two distinct regions. Above 1 bar, $K_{zz}$ decreases exponentially with pressure, because the vertical velocity decreases with pressure. Between 1 and 5 bar, the presence of a detached convective zone induces more vigorous vertical motions, enhances vertical mixing, and consequently increases $K_{zz}$.

We construct approximate `average' parametrisations of the $K_{zz}$ profiles derived from the flux-gradient relationships I and II, as well as from the mixing-length theory (shown as thick black and grey lines in Fig.~\ref{fig:Kzz_profile}). These average $K_{zz}$ profiles are then implemented into the 1D chemical kinetics–transport model \texttt{ATMO} \citep{tremblin_fingering_2015, tremblin2016cloudless, drummond_effects_2016} for validation. 
To enable a direct comparison, we use the same chemical network of \citet{venot2019reduced} as that used in the UM.
For consistency, we adopt the global-mean temperature profile from the 3D synchronous simulation in the 1D model. The metallicity and C/O ratio are set to 180 times solar and solar (0.55), respectively, in agreement with the 3D setup. Each 1D simulation is integrated for 5100 days, matching the duration of the 3D model.

The resulting chemical abundance profiles from the 1D model are compared with those from the 3D simulations in Fig.~\ref{fig:3D_vs_1D}. As shown in Fig.~\ref{fig:3D_vs_1D}a, when adopting the average $K_{zz}$ profile derived from flux–gradient relationship I, the vertical mixing is relatively weak, limiting the upward transport of CO and CO$_2$ from the deep atmosphere to the photosphere. Consequently, the 1D simulations yield lower CO and CO$_2$ abundances in the upper atmosphere compared to the 3D simulation. On the other hand, using the average $K{zz}$ profile derived from the mixing length theory results in overly strong mixing, causing CO, CO$_2$, and NH$_3$ from the deep atmosphere to be transported excessively into the upper atmosphere.
However, when applying the average $K_{zz}$ profile derived from flux-gradient relationship II, the abundances of CO, CO$_2$, NH$_3$, CH$_4$, and H$_2$O in the 1D simulation agree well with those from the 3D simulation. In addition, this 1D simulation successfully reproduces the vertical structure of the chemical species observed in the 3D model, and the quench pressures also show good agreement between the two approaches.
This suggests that the $K_{zz}$ profile derived from the flux-gradient relationship II can serve as an appropriate equivalent $K_{zz}$ profile for use in 1D models of K2-18b, or of exoplanets with similar physical properties, namely an equilibrium temperature of $\sim$300~K and a metallicity of $\sim$180 times solar. It is parametrised as:
\begin{equation}
\label{Kzz_param}
    K_{zz}~(\text{cm}^{2}~\text{s}^{-1})= 
\begin{cases}
    5\times10^4\cdot(\frac{1~\text{bar}}{P_\mathrm{bar}})^{0.7}  & P< 1\,\,\text{bar}\\
    4\times10^5 & 1\,\,\text{bar}\leq P \leq 10\,\,\text{bar}.
\end{cases}
\end{equation}
We note that this parametrisation implicitly assumes that 3D vertical transport can be approximated as a 1D diffusive process, which is most valid in stratified atmospheres where large-scale circulation dominates, and horizontal perturbations are small relative to vertical ones, as discussed above. As it is derived from 3D simulations of K2-18b specifically, therefore most applicable to planets with atmospheric conditions analogous to those of K2-18b.

We note that although the 1D model using the $K_{zz}$ estimated from the flux-gradient relationship II reproduces the CO and CO$_2$ abundances in good agreement with the 3D simulation, the NH$_3$ abundance shows less consistency. This discrepancy may arise because the $K_{zz}$ profile was estimated using an initially uniformly distributed passive tracer with a deep atmospheric source, making it more representative of species like CO and CO$_2$, which also originate from deeper layers. In contrast, NH$_3$ is sourced from the upper atmosphere and exhibits larger initial horizontal contrast and thus may experience a different vertical mixing behaviour. This highlights the challenge of using a single $K{zz}$ profile to accurately represent the transport of multiple chemical species with different source regions.

In Fig.~\ref{fig:Kzz_profile}c, we compare the $K_{zz}$ profile estimated from our simulations with that proposed by \citet[][equation 1]{moses2022chemical} for exo-Neptunes: $K_{zz} = 9.77 \times 10^4 \cdot (1~\mathrm{bar}/P_\mathrm{bar})^{0.5}$~cm$^2$\,s$^{-1}$, which is scaled with an atmospheric scale height of 22.2~km at 0.001~bar and an effective temperature ($\sqrt{2}$ times the equilibrium temperature) of 394.1~K assuming zero albedo for K2-18b. The comparison indicates that the analytical $K_{zz}$ profile from \citet{moses2022chemical} provides a comparable estimation in the radiative convective zone, but cannot capture the stronger vertical mixing at the detached convective zone, suggesting that this scaling relation requires further validation through comprehensive 3D modelling.

\begin{figure*}
\centering
\includegraphics[width=2\columnwidth]{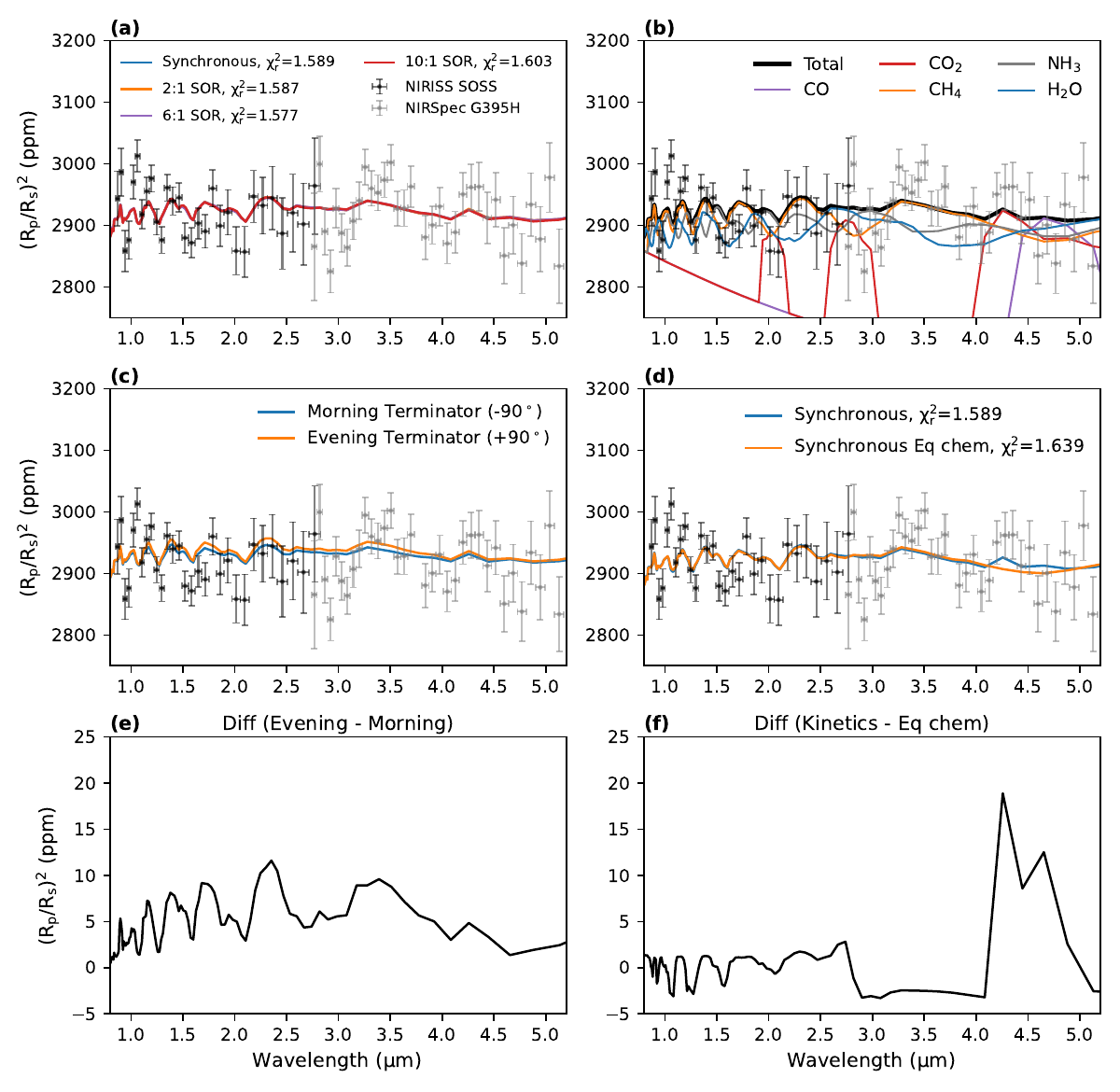}
\caption{K2-18b synthetic transmission spectra predicted by the GCM simulations (coloured lines) compared to JWST NIRISS and NIRSpec data (black and grey error bars) from fig. 3 in \citet{madhusudhan_carbon-bearing_2023}.
Panel (a) shows the full (morning plus evening terminators) transmission spectra from chemical kinetics–transport simulations, assuming K2-18b as a synchronous rotator (blue), or has a 2:1 SOR (orange), 6:1 SOR (purple), and 10:1 SOR (red). The spectra from different simulations closely overlay.
Panel (b) shows the contributions from the dominant chemical species, CO, CO$_2$, NH$_3$, CH$_4$, and H$_2$O, with the synchronous chemical kinetics–transport simulation as an example. Panel (c) depicts the morning (blue) and evening (orange) transmission spectra in the synchronous chemical kinetics–transport simulation.
Panel (d) shows the effects of transport-induced disequilibrium chemistry on observation, in which the blue line shows the transmission spectrum from synchronous chemical kinetics–transport simulations, while the orange line shows the spectrum under chemical equilibrium, calculated using the same 3D temperature field as the synchronous chemical kinetics–transport simulations.
The $\chi_r^2$ values shown in panels (a) and (d) are calculated with 64 degrees of freedom with one offset between NIRISS and NIRSpec data. Panels (e) and (f) depict the difference between the evening and morning transmission spectra and the difference between the spectra generated from chemical kinetics–transport and chemical equilibrium simulations.}
\label{fig:STS_3D}
\end{figure*}

\subsection{Synthetic Transmission Spectra} \label{subsec:observation}

In this section, we present the synthetic transmission spectra generated from the GCM simulations and compare them to the JWST observations reported by \citet{madhusudhan_carbon-bearing_2023} and \citet{Hu_K2-18b_2025}. We highlight the differences between the morning and evening transmission spectra and examine the impact of transport-induced disequilibrium chemistry on the transmission spectra of K2-18b.

We generate the transmission spectra following the method described in \citet{Lines_trans_spectra_2018} and updated by \citet{christie_impact_2021}. 

We first compute the full-limb transmission spectra, which include contributions from both the morning and evening terminators, and compare them with the JWST observations presented in fig.~3 of \citet{madhusudhan_carbon-bearing_2023} (shown in Fig.~\ref{fig:STS_3D}a). 
Overall, the synthetic spectra from GCM simulations agree well with the JWST observations, with a $\chi_r^2$ value of around 1.6, 
suggesting that the gas-rich mini-Neptune scenario is a reasonable explanation for the interior structure of K2-18b. The synthetic spectra from different simulations remain almost invariant with each other. This is due to the air temperature profiles and molecular abundances being consistent across simulations (see Fig.~\ref{fig:mole_profile}). 

The contributions from the dominant molecules CO, CO$_2$, NH$_3$, CH$_4$, and H$_2$O are shown in Fig.~\ref{fig:STS_3D}b. CH$_4$ contributes to the absorption features at wavelengths shorter than 2.5~$\mu$m and at 3.3~$\mu$m. CO$_2$ mainly contributes to the feature at 4.3~$\mu$m, and the small bump at 4.6~$\mu$m is produced by CO. Compared with the observations, our transmission spectra may over-predict the NH$_3$ and CO absorption features at 2.9 and 4.6~$\mu$m.

Compared with the retrieved abundances reported in \citet{madhusudhan_carbon-bearing_2023}, our 3D model yields higher abundances of NH$_3$, H$_2$O, and CO, while the abundances of CO$_2$ ($10^{-3}$) and CH$_4$ (0.1) are in broad agreement. For their one-offset retrieval, the reported 1$\sigma$ ranges are $0.9\times10^{-3}$–0.026 for CO$_2$ and 0.037–0.07 for CH$_4$. The low NH$_3$ abundance inferred from the data may arise if the planet hosts a deep magma ocean capable of dissolving NH$_3$, which is not included in our model, or if the planet is intrinsically nitrogen-poor \citep{shorttle_distinguishing_2024,Hu_K2-18b_2025}. The elevated CO abundance in our simulation likely results from the absence of photochemistry, which would otherwise dissociate CO. The enhanced H$_2$O abundance likely arises from the omission of H$_2$O condensation in our model.

The transmission spectra calculated solely with the morning or evening terminator in the synchronous simulation are shown in Fig.~\ref{fig:STS_3D}c. The transmission spectrum of the evening terminator exhibits larger absorption features because it is warmer than the morning terminator due to the eastward heat transport by the eastward winds above 0.1 bar. Spectral differences between the two terminators are most apparent at CH$_4$ absorption bands at 2.3, 3.3, and 1.8~$\mu$m, corresponding to a 12, 9, and 9~ppm difference in transit depth (Fig.~\ref{fig:STS_3D}e). The spectral differences caused by the CO$_2$ and CO absorption at 4.3 and 4.6~$\mu$m are much smaller, corresponding to only 5 and 2~ppm difference in transit depth. 

As shown in Fig.~\ref{fig:STS_3D}d, the synthetic spectrum from the chemical kinetics–transport simulation provides a better match to the observational data, particularly capturing the CO$_2$ absorption feature at 4.3~$\mu$m. This feature corresponds to a transit depth difference of 21~ppm, compared to the spectrum from the chemical equilibrium simulation (Fig.~\ref{fig:STS_3D}f). The CO absorption feature at 4.6~$\mu$m corresponds to a transit depth difference of 12~ppm between the chemical kinetics–transport and chemical equilibrium simulations.
As previously discussed, these features result from the vertical transport of CO$_2$ and CO from the deep atmosphere driven by atmospheric circulation. This highlights the crucial role of vertical mixing in shaping observable spectral characteristics. However, all of these differences mentioned above remain within the current 1$\sigma$ observational uncertainties.

\begin{figure*}
\centering
\includegraphics[width=2\columnwidth]{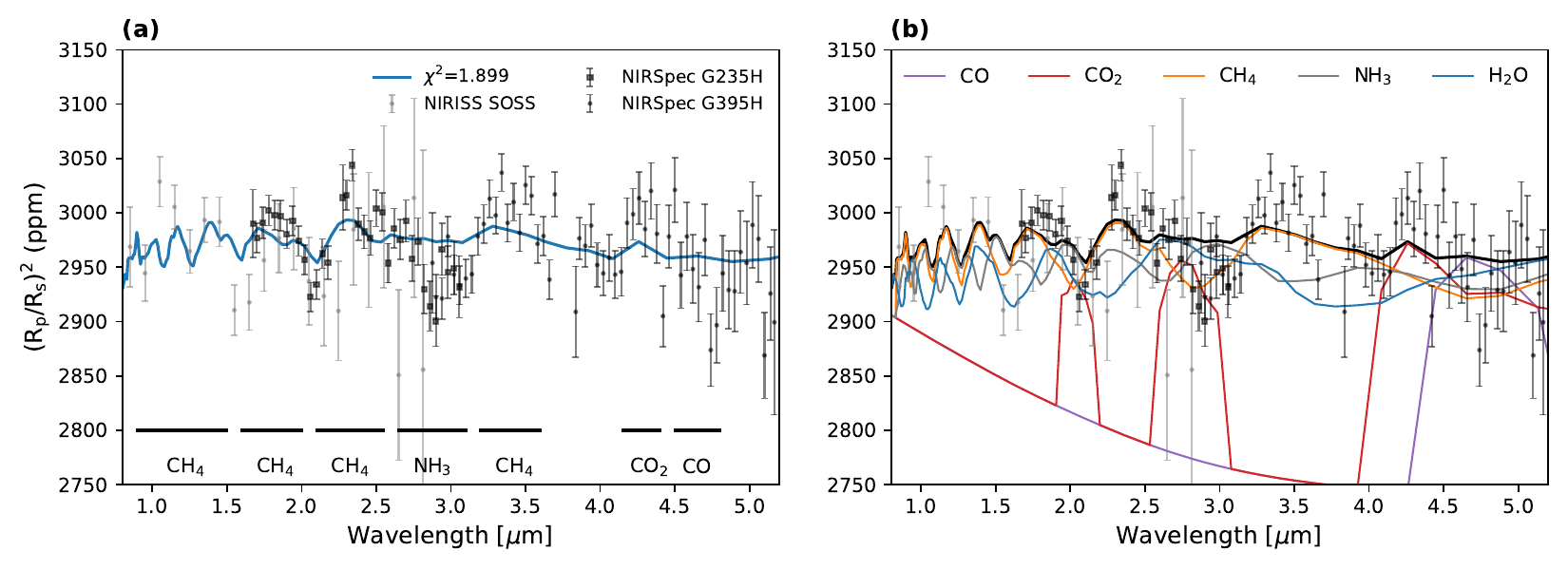}
\caption{K2-18b synthetic transmission spectrum predicted by the GCM simulation compared to JWST NIRISS and NIRSpec data (grey and black error bars) from fig. 3 in \citet{Hu_K2-18b_2025}.
Panel (a) shows the full (morning plus evening terminators) transmission spectrum from the synchronous chemical kinetics–transport simulation as an example.
Panel (b) shows the contributions from the dominant chemical species, CO, CO$_2$, NH$_3$, CH$_4$, and H$_2$O.}
\label{fig:STS_3D_Hu}
\end{figure*}

We note that \citet{Hu_K2-18b_2025} released a new JWST near-infrared transmission spectrum of K2-18b while we were preparing this manuscript. We compared our synthetic transmission spectra with their observations. Our spectra broadly agree with their results and successfully reproduce the CO$_2$ and CH$_4$ absorption features. However, our spectra may over-predict the NH$_3$ and CO absorption features near 2.9 and 4.6~$\mu$m (Fig.~\ref{fig:STS_3D_Hu}).

\section{Conclusions and Discussion}\label{sec:conclusions}

In this study, we present the 3D perspective on transport-induced disequilibrium chemistry on the temperate mini-Neptune K2-18b using a self-consistent chemical-radiative-dynamic GCM. We assume K2-18b has 180 times solar metallicity and consider it as both a synchronous rotator and an asynchronous rotator, exploring the spin-orbit resonance (SOR) of 2:1, 6:1, and 10:1. 

We find that vertical mixing significantly affects the molecular abundances in the photosphere. The quench levels of the dominant species: CO, CO$_2$, CH$_4$, and H$_2$O are located around 7~bar, while NH$_3$ quenches deeper, at approximately 15~bar. Among these species, CO and CO$_2$ exhibit the most significant enhancement under transport-induced disequilibrium chemistry: their mole fractions increase from below 10$^{-15}$ in chemical equilibrium to approximately 10$^{-3}$ due to vertical mixing that transports them upward from deeper, more abundant layers. In contrast, the mole fractions of NH$_3$, CH$_4$, and H$_2$O decrease slightly compared to chemical equilibrium, as these species are more stable under low-temperature conditions. Specifically, their mole fractions drop from 0.025, 0.11, and 0.19 to 0.01, 0.10, and 0.18 for NH$_3$, CH$_4$, and H$_2$O, respectively.

Strong horizontal winds efficiently homogenise chemical abundances in the photosphere, especially in the zonal direction. Although some meridional gradients remain, they are relatively small. The meridional contrasts of chemical species are the combined results of atmospheric dynamics and chemical calculation, highlighting the importance of self-consistent chemical-radiative-dynamic modelling in studying the transport-induced chemistry on temperate sub-Neptunes.
Furthermore, we find that global-mean molecular abundances in the photosphere are largely consistent across simulations with different rotation periods, indicating that the rotation rate has a limited effect on global-mean vertical mixing.

We estimate 1D equivalent eddy diffusion coefficient ($K_{zz}$) profiles from 3D GCM simulations using both the flux-gradient relationship and mixing-length theory. Our results show that $K_{zz}$ profiles estimated from the two methods are similar in shape. At pressures between 1 and 5 bar, the presence of a detached convective zone enhances vertical mixing, resulting in an increased $K_{zz}$. At the radiative-convective zone above 1 bar, $K_{zz}$ decreases with increased pressure due to the decreasing vertical velocity. The $K_{zz}$ values derived from mixing-length theory are one to two orders of magnitude larger than those estimated from the flux-gradient relationship, consistent with previous studies. We also find that vertical mixing in the GCM exhibits significant temporal variability. 

By implementing the estimated $K_{zz}$ profile into the 1D chemical kinetics–transport model \texttt{ATMO}, which uses the same chemical network as the 3D model, we find that the $K_{zz}$ profile estimated based on the mixing-length theory (equation \ref{eq:mixing_length_theory}) induced too strong vertical mixing strength, while the flux-gradient relationship I (equation \ref{eq:flux-gradient I}) slightly underestimates the strength of vertical mixing. In contrast, the $K_{zz}$ profile estimated from flux-gradient relationship II (equation \ref{eq:flux-gradient II}) successfully reproduces molecular abundances and vertical profiles in the photosphere that closely match the 3D results, providing an appropriate equivalent $K_{zz}$ profile that can be further used in 1D models. It is parametrised as equation \ref{Kzz_param}.

The development of the detached convective zone induced by CO$_2$ and CH$_4$ absorption implies that atmospheric composition may significantly impact vertical mixing. These results further emphasise the importance of a non-grey treatment of radiative processes and self-consistent chemical-radiative-dynamic calculations in studying transport-induced chemistry on gas-rich exoplanets.

We then compare the synthetic transmission spectra from the 3D simulations with the JWST observations of K2-18b in \citet{madhusudhan_carbon-bearing_2023} and \citet{Hu_K2-18b_2025}. We find that our 3D simulations provide comparable fits to the observations, suggesting that the gas-rich mini-Neptune scenario is a reasonable explanation for the interior structure of K2-18b. The transmission spectra are almost invariant between different rotation periods because the temperature profiles and the molecular abundances agree well among different simulations. 

The evening transmission spectrum exhibits stronger absorption features than the morning spectrum, primarily due to the higher temperature on the evening terminator, a consequence of eastward heat transport by upper-atmosphere eastward winds. The largest difference between evening and morning spectra appears at the 2.3~$\mu$m CH$_4$ feature, with a transit depth difference of 12~ppm.

Furthermore, we find that the spectrum from transport-induced disequilibrium chemistry provides a better agreement with the JWST observations compared to the spectrum generated from chemical equilibrium, as it better captures the CO$_2$ absorption feature at 4.3~$\mu$m. This corresponds to a 21~ppm transit depth difference between the chemical kinetics–transport and chemical equilibrium simulations, highlighting the important role of atmospheric vertical transport in shaping the observation features.

Our results have two broader implications for the atmospheric characterisation of temperate gas-rich planets. First, vertical mixing plays a crucial role in determining the photospheric molecular abundances of such planets. Atmospheric retrievals and modellings of temperate mini-Neptunes, including promising targets such as TOI-270d \citep{mikal-evans_2023} and LP~791-18c \citep{peterson_2023}, should therefore adopt disequilibrium chemistry assumptions and a physically motivated $K_{zz}$ profile, to avoid biased estimates of atmospheric metallicity and molecular abundances. Second, the rotation period has a limited influence on both the synthetic transmission spectra and the globally averaged chemical abundance profiles, suggesting that the uncertainty in planetary rotation period of such relatively long-period planets may not need to be explicitly considered in 1D retrieval frameworks at the current level of observational precision.

As with any modelling study, our simulations necessarily involve some simplifications. Photochemistry is not included in this study. While photochemistry may play a significant role in the photosphere of K2-18b, considering the possible extremely long convergence time for 3D simulation of sub-Neptunes \citep{wang2020extremely}, fully coupled 3D photochemical-radiative-hydrodynamic modelling within a GCM might require significant computational resources, making it less practical. Further works on photochemical calculations in GCM can consider coupling a much computationally lighter chemical model \citep[e.g.][]{Tsai_2022,Lee_2023}, or implement schemes to accelerate the convergence time to equilibrium by artificially reducing the radiative timescale in the deep atmospheric layers \citep[e.g.][]{turbet_daynight_2021}. Other efficient and practical approaches to investigate photochemical kinetics on temperate sub-Neptunes include using the equivalent $K_{zz}$ profile derived from our study (equation \ref{Kzz_param}) in the 1D models or adopting the wind speed fields from our GCM simulations in 2D photochemical models \citep[e.g.][]{tsai_inferring_2021,Baeyens_2022}.  

Condensation of chemical species is neglected in the simulations. However, the intersection between the water condensation curves and the air temperature profiles shown in figs 2a--d in \citetalias{Liu_2025_PartI} suggests that water clouds may form at the photosphere. Although \citet{charnay2021formation} suggested that these clouds have a relatively small impact on transmission spectra, the condensation of H$_2$O can suppress convection in hydrogen-dominated atmospheres, a phenomenon commonly referred to as `convective inhibition' \citep{guillot1996giant,Leconte_2017,Leconte_2024,Gao_hycean}. This convective inhibition can suppress vertical transport, reduce $K_{zz}$, and potentially affect molecular abundances in the photosphere.  
Additionally, a recent study by \citet{Habib_3D_2024} reported that convective inhibition could lead to the depletion of water vapour in the upper atmosphere, preventing its detection. While only a small region of the atmosphere in our simulations has the potential to undergo convective inhibition, this remains an important process to consider in future studies, especially for cooler planets. This also highlights the complex nature of vertical mixing on temperate sub-Neptunes.

\section*{Acknowledgements}
We thank the anonymous referee for comments that improved the quality of this manuscript. We are grateful to Nathan Mayne for providing access to the UM and for his assistance with running the model, and to Eva-Maria Ahrer and Feng Ding for helpful discussions. We also thank Mei Ting Mak for her support with the transmission spectrum calculations, and Shang-Min Tsai for raising a question at ExoClimes~VI that helped us identify and resolve an issue in our 1D simulations.
This work used the Max Planck Society's Viper High-Performance Computing system. 
This work also used the DiRAC Complexity system, operated by the University of Leicester IT Services, which forms part of the STFC DiRAC HPC Facility and the University of Exeter Supercomputer ISCA. We note that the production runs used in this paper required about 2.3 million CPU hours. The use due to testing and analysis is about 0.5 million CPU hours.
J. Liu is partly supported by the Overseas Study Program for Graduate Students of the China Scholarship Council. J.Y. is supported by the National Science Foundation of China (NSFC) under grant no. 42275134.

The analysis and visualisation of the simulation data made use of the following {\sc python} packages: {\sc aeolus} \citep{sergeev2023aeolus}, {\sc iris} \citep{scitools_iris_2023}, {\sc matplotlib} \citep{hunter_maplotlib}, and {\sc numpy} \citep{harris2020array}.

\section*{Data Availability}

The simulation data used in this study are available from the corresponding author upon reasonable request.



\bibliographystyle{mnras}
\bibliography{example} 


\appendix

\section{Thermal structure and wind structure}\label{sec:figures_PartI}

\begin{figure*}
\includegraphics[width=1.9\columnwidth]{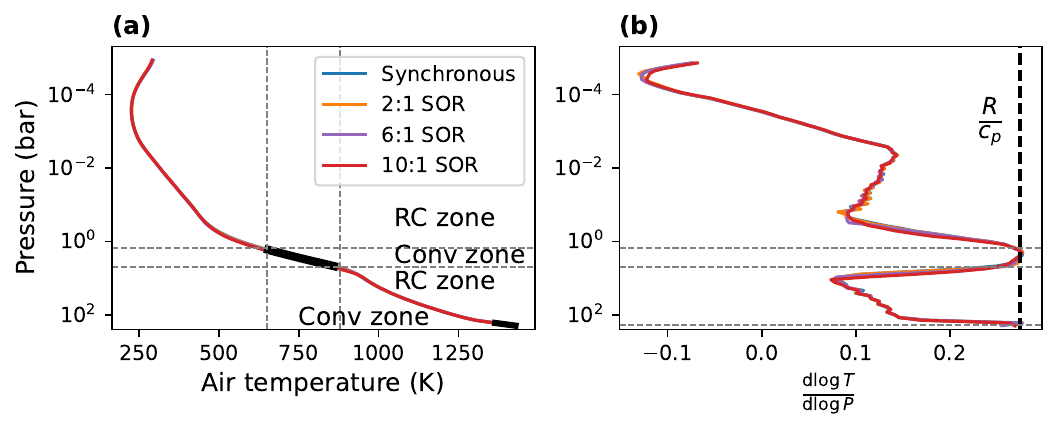}
\caption{Global mean air temperature vertical profiles (a) and global mean air temperature lapse rate profiles (b) in all the simulations, reproduced from fig.~2 of \citetalias{Liu_2025_PartI} and included here for reference. Different coloured lines represent results from simulations with different rotation periods. Coloured lines overlap, indicating that global mean air temperature profiles are consistent across different rotation periods.
The dashed thick black line in panel (b) is the dry adiabatic temperature lapse rate, and its intersection with the local temperature lapse rate indicates the appearance of convective zones. The thin grey dashed lines indicate the boundary between the radiative-convective (RC) zones and the convective (Conv) zone.}
\label{fig:thermal_structure}
\end{figure*}

\begin{figure*}
\includegraphics[width=2\columnwidth]{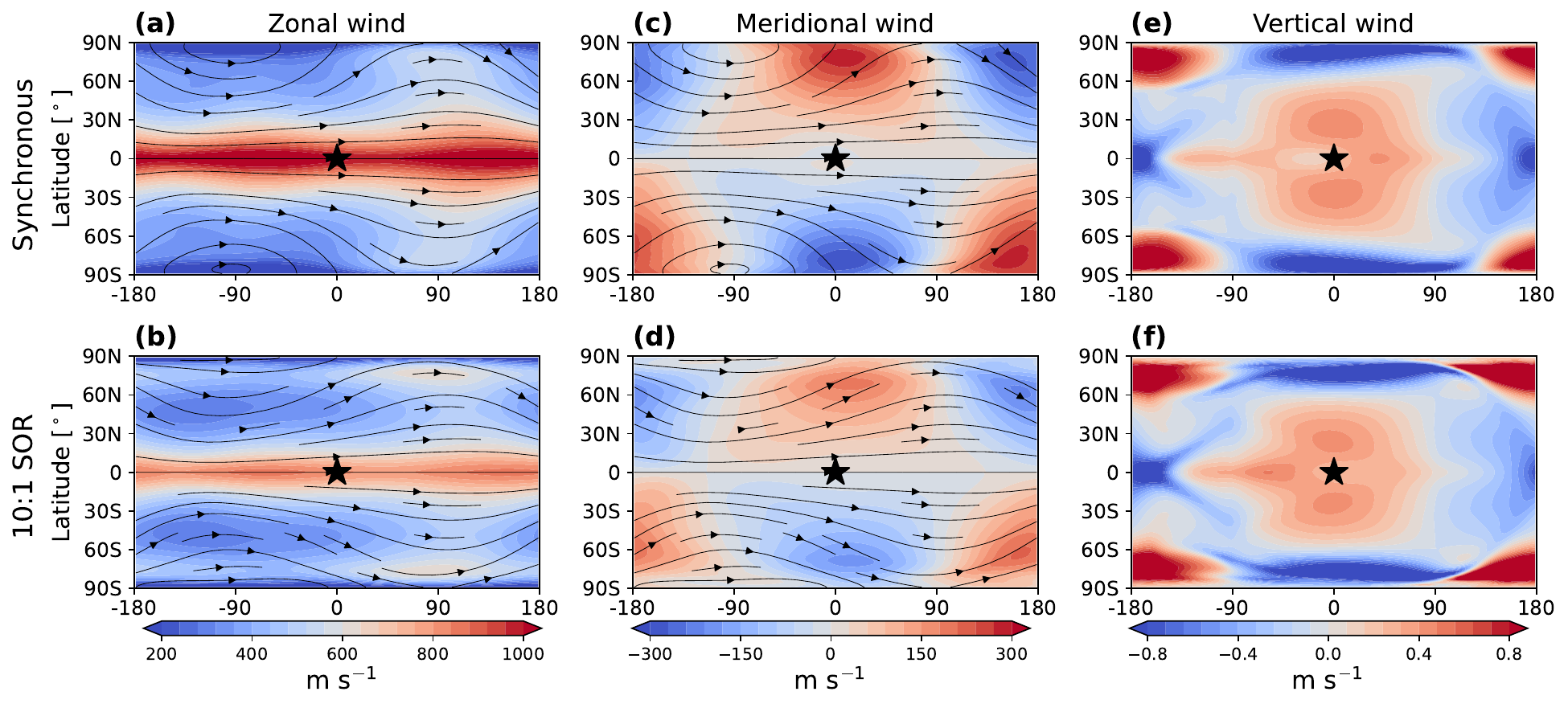}
\caption{Wind structures for the synchronous and 10:1 SOR simulations, reproduced from fig.~5 of \citetalias{Liu_2025_PartI} and included here for reference. Columns from left to right show the horizontal distribution of zonal wind, meridional wind, and vertical wind at 0.001 bar. Streamlines indicate the direction of the horizontal flow. Results are shown in the heliocentric frame to more clearly illustrate the wind patterns. Black star-shaped markers indicate the location of the substellar point. The first and second rows show results from the synchronous and 10:1 SOR simulations as examples, with the wind patterns in other simulations being similar.}
\label{fig:wind_pattern}
\end{figure*}

We show the thermal structures and wind structures of the simulations for reference in the context of the present study.
Fig.~\ref{fig:thermal_structure} shows the global-mean thermal structure for all simulations. Fig.~\ref{fig:wind_pattern} shows the wind patterns for synchronous and 10:1 SOR simulations as examples, with the wind patterns in other simulations being similar.

\section{Additional Information}

\begin{figure*}
\centering
\includegraphics[width=1.5\columnwidth]{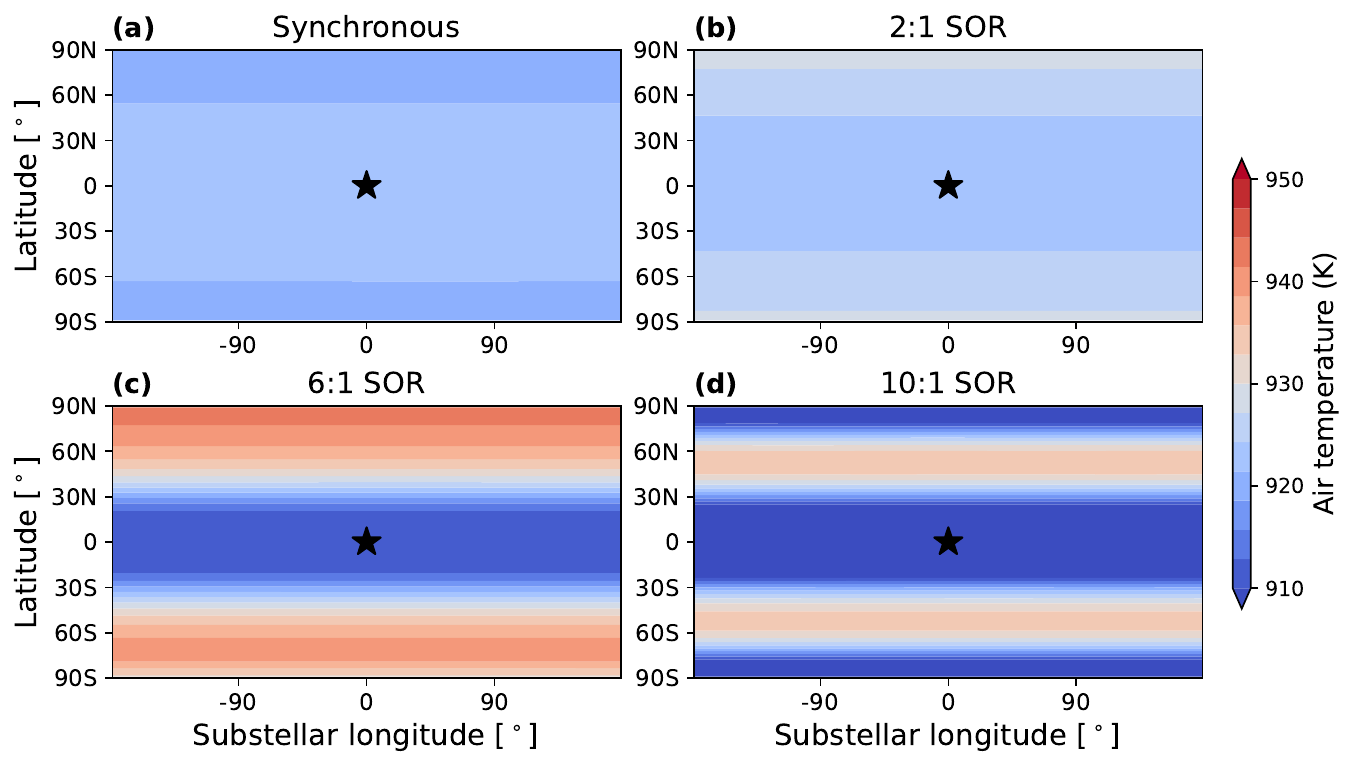}
\caption{Horizontal distribution of air temperature at 7 bar in the heliocentric frame. Panels from (a) to (d) are results from synchronous, 2:1 SOR, 6:1 SOR, and 10:1 SOR simulations, respectively.
}
\label{fig:tair_7bar}
\end{figure*}

\begin{figure*}
\centering
\includegraphics[width=1.5\columnwidth]{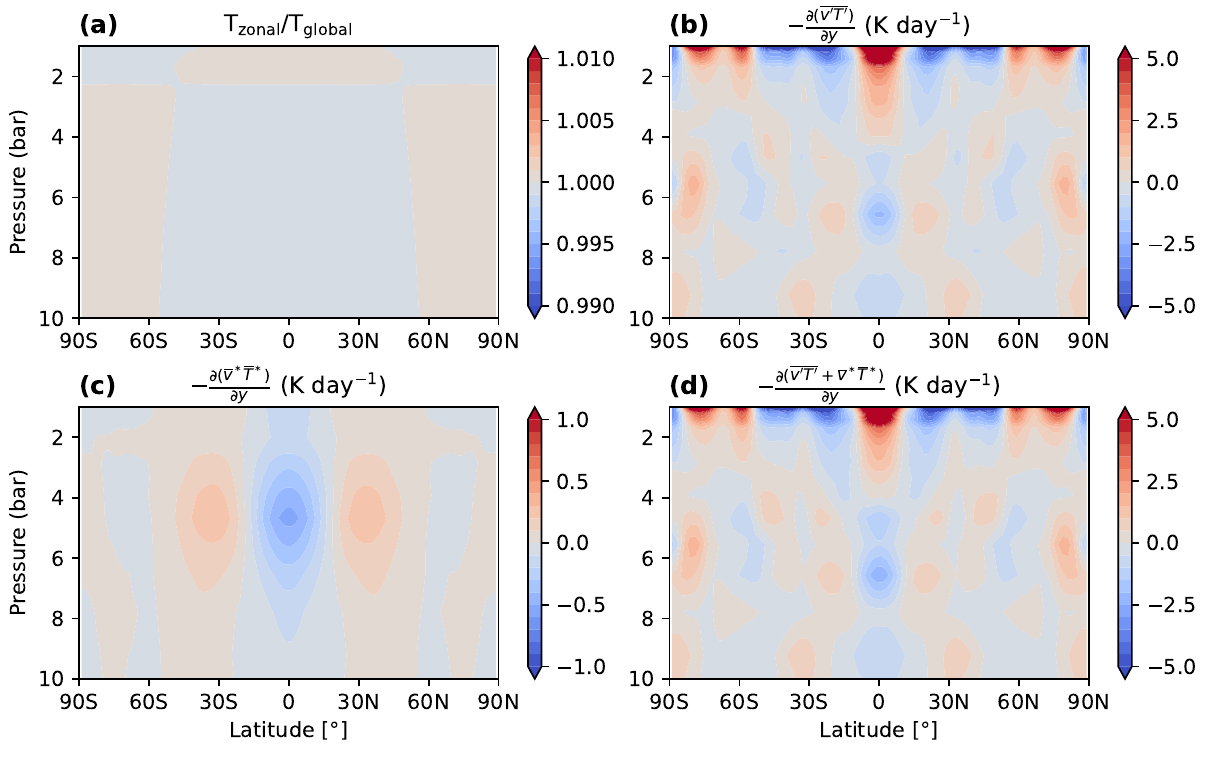}
\caption{Pressure versus latitude distribution of air temperature contrast (a), transient eddy heat transport (b), stationary eddy heat transport (c), and total eddy heat transport (d) for the 6:1 SOR simulation. Results are derived from simulated data over the first 100 d, in which the horizontal temperature contrast begins to form, with an output interval of 10 d.
}
\label{fig:eddy_heat}
\end{figure*}

Fig. \ref{fig:tair_7bar} shows the horizontal distribution of air temperature at 7 bar. In the 2:1 SOR, 6:1 SOR, and 10:1 SOR simulations, the polar regions are much warmer than the equatorial regions, leading to enhanced CO and CO$_2$ abundances at high latitudes at this level (Fig. \ref{fig:mole_quench}). We argue that this reversed temperature contrast might result from the eddy heat convergence at the polar regions (Fig. \ref{fig:eddy_heat}), transporting heat from low latitudes to high latitudes. When ignoring the horizontal perturbation of air density at 1 to 10 bar for simplicity, the eddy heat flux convergence is attributed to the increase of air temperature: $\frac{\partial T}{\partial t} \sim -\frac{\partial (\overline{v'T'}+\overline{v}^*\overline{T}^*)}{\partial y}$ \citep{hartmann_atmospheric_2007}. However, due to the coarse vertical resolution between 5 and 10 bar ($\sim$1 bar), these temperature contrasts could also result from interpolation errors, random noise, or unresolved dynamical processes.

\begin{figure*}
\centering
\includegraphics[width=1\columnwidth]{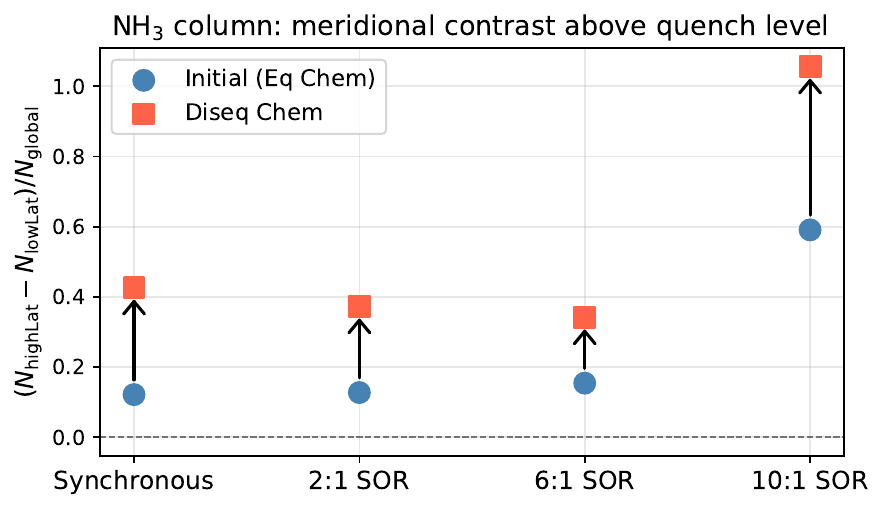}
\caption{Column-integrated NH$_3$ mole fraction contrast between high- ($>75^\circ$) and low- ($<15^\circ$) latitude regions above the quench level (15 bar). The vertical axis shows the difference between the high- and low-latitude column abundances normalised by the global mean. Blue circles and orange squares denote initial (under chemical equilibrium) and transport-induced disequilibrium states, respectively. Arrows indicate the direction and magnitude of the shift induced by disequilibrium chemistry. Positive values indicate that NH$_3$ is more abundant at high latitudes. The meridional contrast between high- and low-latitude regions increases under disequilibrium chemistry across all rotation states.}
\label{fig:NH3_change}
\end{figure*}

Fig.~\ref{fig:NH3_change} shows the column-integrated NH$_3$ meridional contrast above the quench level (15 bar) in the initial thermochemical equilibrium state and under transport-induced disequilibrium chemistry. The meridional contrast increases under disequilibrium chemistry across all rotation states.

\bsp	
\label{lastpage}
\end{document}